\documentclass[usenatbib]{mnras}
\usepackage{amsmath,amsfonts,amssymb}
\usepackage{textcomp,gensymb}
\usepackage[T1]{fontenc}
\usepackage{aecompl}
\usepackage{ulem}
\usepackage[pdftex]{graphicx}
\usepackage{aas_macros}
\usepackage[usenames]{xcolor}
\usepackage{color}
\usepackage{times}
\usepackage{subfig}
\usepackage{stfloats}
\usepackage[utf8]{inputenc}
\usepackage{hyperref}
\usepackage{multirow}

\newcommand{\ans}[1]{\textcolor{black}{#1}}
\newcommand{\ANS}[1]{\textcolor{black}{#1}}

\newcommand{\hi}[1]{\textcolor{black}{#1}}

\title[The circumbinary disc of HD~169142]{The protoplanetary disc around HD 169142: circumstellar or circumbinary?}

\author[Poblete et al.]{\parbox{\textwidth}{P. P. Poblete,$^{1,2}$ N. Cuello,$^{3}$ S. Pérez,$^{4, 5}$ S. Marino,$^{6,7}$, J. Calcino,$^{8, 9}$ E. Macías,$^{10,11}$ Á. Ribas,$^{11}$ A. Zurlo,$^{12,13}$ J. Cuadra,$^{14,2}$ M. Montesinos,$^{15,2}$ S. Zúñiga-Fernández,$^{2,11,16}$ A. Bayo,$^{16,2}$ C. Pinte,$^{17,3}$ F. M\'enard$,^{3}$ and D.~J. Price$^{17}$}\vspace{0.15cm}\\
$^{1}$Astrophysikalisches Institut, Friedrich-Schiller-Universität Jena, Schillergäßchen 2–3, 07745 Jena, Germany\\
$^{2}$N\'ucleo Milenio de Formaci\'on Planetaria (NPF), Chile, \\
$^{3}$Univ. Grenoble Alpes, CNRS, IPAG (UMR 5274), F-38000 Grenoble, France\\
$^{4}$Departamento de F\'isica, Universidad de Santiago de Chile. Avenida Ecuador 3493, Estaci\'on Central, Santiago, Chile\\
$^{5}$Center for Interdisciplinary Research in Astrophysics and Space Exploration (CIRAS), Universidad de Santiago de Chile, Estaci\'on Central, Chile\\ 
$^{6}$Jesus College, University of Cambridge, Jesus Lane, Cambridge CB5 8BL, UK\\
$^{7}$Institute of Astronomy, University of Cambridge, Madingley Road, Cambridge CB3 0HA, UK\\
$^{8}$Theoretical Division, Los Alamos National Laboratory, Los Alamos, NM 87545, USA \\
$^{9}$School of Mathematics and Physics, The University of Queensland, QLD 4072, Australia \\
$^{10}$Joint ALMA Observatory, Alonso de C\'ordova 3107, Vitacura, Casilla 19001, Santiago de Chile, Chile \\
$^{11}$European Southern Observatory, Alonso de C\'ordova 3107, Vitacura, Casilla 19001, Santiago de Chile, Chile \\
$^{12}$N\'ucleo de Astronom\'ia, Facultad de Ingenier\'ia y Ciencias, Universidad Diego Portales, Av. Ejercito 441, Santiago, Chile\\
$^{13}$Escuela de Ingenier\'ia Industrial, Facultad de Ingenier\'ia y Ciencias, Universidad Diego Portales, Av. Ejercito 441, Santiago, Chile \\
$^{14}$Departamento de Ciencias, Facultad de Artes Liberales, Universidad Adolfo Ib\'a\~nez, Avenida Padre Hurtado 750, Vi\~na del Mar, Chile
\\
$^{15}$ Escuela de Ciencias, Universidad Vi\~na del Mar, Vi\~na del Mar, Chile \\
$^{16}$ Instituto de F\'isica y Astronom\'ia, Facultad de Ciencias, Universidad de Valpara\'iso, Av. Gran Breta\~na 1111, Valpara\'iso, Chile\\
$^{17}$School of Physics and Astronomy, Monash University, Clayton VIC 3800, Australia
}

\begin{document}
\date{Accepted 2021 November 25. Received 2021 November 25; in original form 2021 June 9}

\pagerange{\pageref{firstpage}--\pageref{lastpage}} \pubyear{2020}

\maketitle

\label{firstpage}

\begin{abstract}
Stellar binaries represent a substantial fraction 
of stellar systems, especially among young stellar objects. Accordingly, binaries play an important role in setting the architecture of a large number of protoplanetary discs. Binaries in coplanar and polar orientations with respect to the circumbinary disc are stable configurations and could induce non-axisymmetric structures in the dust and gas distributions.
In this work, we suggest that the structures shown in the central region of the protoplanetary disc HD~169142 are produced by the presence of an inner stellar binary and a circumbinary (P-type) planet. We find that a companion with a mass-ratio of $0.1$, semi-major axis of \hi{$9.9$~au}, eccentricity of \hi{0.2}, and inclination of $90\degree$, together with a \hi{2~$M_{\rm J}$ coplanar planet on a circular orbit at 45~au} reproduce the structures at the innermost ring observed at 1.3 mm and the shape of spiral features in scattered light observations. The model predicts changes in the disc's dust structure, and star's astrometric parameters, which would allow testing its veracity by monitoring this system over the next 20 years.
\end{abstract}

\begin{keywords}
protoplanetary discs --- hydrodynamics --- methods: numerical ---  binaries (including multiple): close --- planet-disc interactions --- stars: individual: HD~169142
\end{keywords}


\section{Introduction}
\label{sec:intro}

One of the most outstanding features in some protoplanetary discs is the presence of a cleared-out central cavity, so-called transitional discs or TD \citep{Strom+1989}. In recent years, the dynamical origin for such central cavities of TDs has been gaining traction. \citep[e.g. ][]{Calvet+2002, Espaillat+2010,Casassus2016, Stolker+2016, vanderMarel+2018, Price+2018, Francis&Marel2020}.
However the physical properties of the companions responsible for these cavities remain unconstrained by observations in all but a few cases \citep[e.g. CoKu Tauri/4, HD~142527, PDS~70,][]{Ireland&Kraus2008,Biller2012, Lacour+2016, Keppler+2018, Haffert2019,Wang+2020}. Companions ranging from planetary to stellar mass, on circular and co-planar \citep[e.g.][]{vanderMarel+2013, Calcino+2019} to inclined and elliptical orbits \citep[e.g.][]{Price+2018, Poblete+2020, Calcino+2020}, have been proposed.

The spiral arms observed in the near-infrared, and structures in the thermal emission of dust grains at (sub-)millimeter wavelengths among other sub-structures on the TDs, have been widely studied. Better understanding these sub-structures, through both observations and numerical modeling is necessary to assess their origin. In this context, \citet{Poblete+2019} \ans{showed} that the presence of dusty clumps embedded in a dust ring can provide information on the mass, inclination, and eccentricity of a companion located inside the dust ring.
Such clumps resemble some of the asymmetric structure seen in the protoplanetary disc around HD~169142 \citep{Macias+2017,Perez+2019}, and thus they could be produced by an inner massive and undetected companion. 

\begin{figure}
\centering
\begin{center}
    \includegraphics[width=0.5\textwidth]{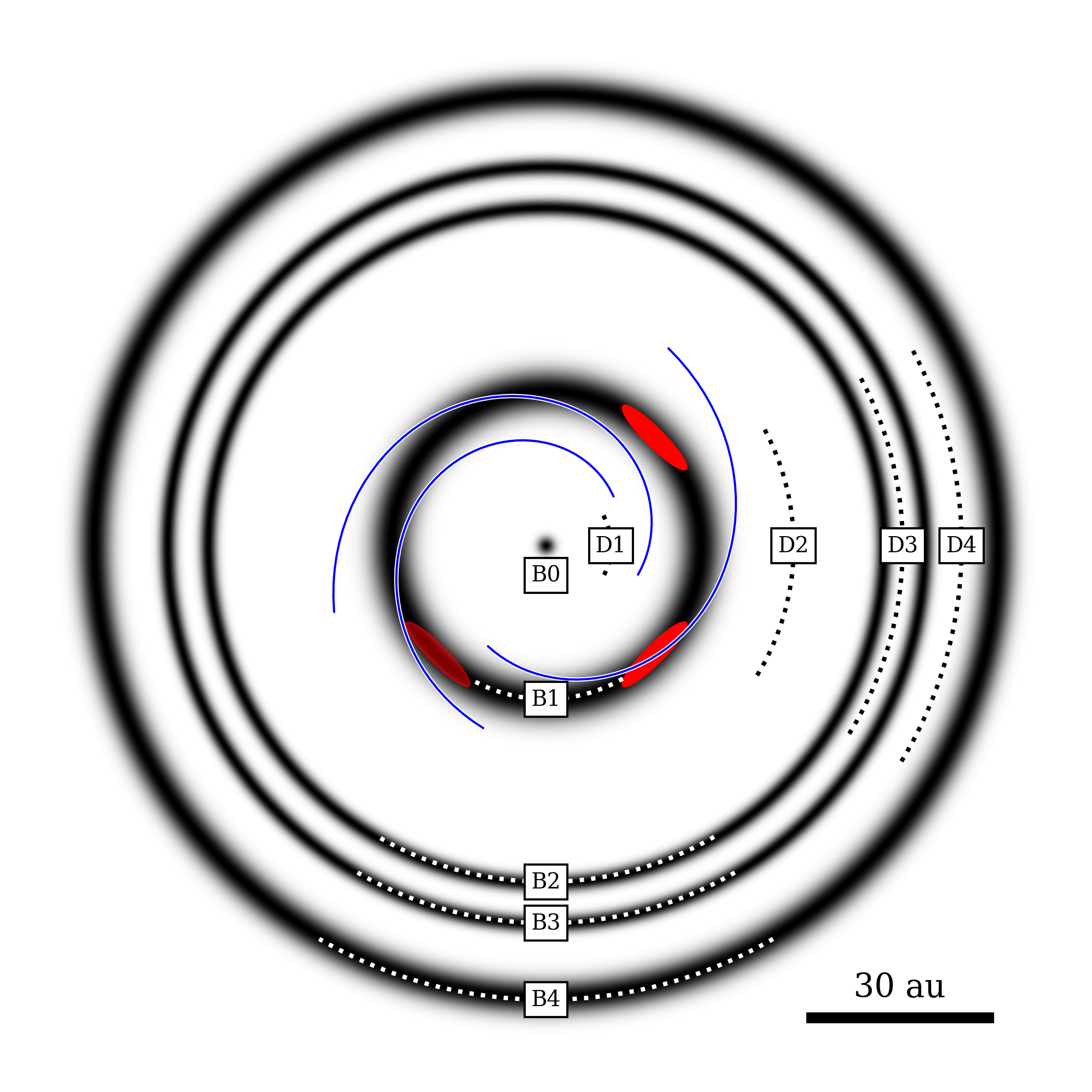}
    \caption{Schematic illustration of the structures observed in the disc of HD~169142. The disc exhibits numerous gaps or dips (D1, D2, D3, and D4), a central misaligned circumstellar disc (B0), and rings (B1, B2, B3, and B4 with center at 26, 57, 64 and 77 au respectively).  
    The B1  ring shows three dust clumps highlighted in red. The clumps appear to be separated by $\sim90\degree$ among them, with one of them less intense than the other two. Besides, three spirals (in blue) appear crossing the region defined by D1, B1 and D2.
    }
    \label{fig:sketch}
\end{center}
\end{figure}

The star HD~169142 has been intensively studied at multiple wavelengths. At a distance of 113.6$\pm0.8$ pc \citep{Gaia2018,Bailer-Jones+2018} this F1 pre-main sequence star \citep{Gray+2017} is estimated to have a mass of 1.65$\pm0.5\ M_{\odot}$ \citep{Carney+2018} and an age of $6_{-3}^{+6}$ Myr \citep{Grady+2007}. The star is surrounded by an inner circumstellar disc with a radius of 2.2$\pm0.6$~au \citep{Perez+2019, Francis&Marel2020} that is misaligned between 10$\degree$--$23\degree$ with respect to an outer disc \citep{,Lazareff+2017,Chen+2018,Francis&Marel2020}. The outer disc is inclined by $\sim12.5\degree$ \citep{Raman+2006,Panic+2008} and extends from 10~au to roughly 120~au; whilst the mm dust continuum emission extends from 26~au to roughly 90~au \citep{Fedele+2017,Perez+2019,Macias+2019}. The outer disc also shows several prominent features such as: a multi-ring structure \citep{Quanz+2013,Monnier+2017,Pohl+2017,Perez+2019}, spirals \citep{Ligi+2018,Bertrang+2018,Gratton+2019}, and dust clump structures at the innermost ring \citep{Osorio+2014,Macias+2017,Macias+2019,Perez+2019}. Figure \ref{fig:sketch} illustrates the disc morphology. Note that the spirals only appear after subtracting an azimuthal average or after filtering by Angular Differential Imaging. \ANS{A detailed explanation of the spiral morphology is given in Section \ref{sec:gas}.}

Mechanisms to explain disc axisymmetrical structures include photo-evaporation \citep{Alexander+2006,Alexander+2014}, dead-zones \citep{Flock+2015,Pinilla+2016}, disc-wind interactions \citep{Suriano+2018}, magneto-hydrodynamic processes \citep{Lesur+2014,Riols&Lesur2018}, dust processes \citep{Pinilla+2012a,Pinilla+2017, Cuello+2016,Gonzalez+2017}, and planet-disc interactions \citep{Lin+1986,Pinilla+2012b,Zhu&Stone2014,Bae+2017,Zhang+2018}. The latter is the most promising to describe the multi-ring structure in HD~169142. Different planetary configurations have been proposed to reproduce the observations, but in general the disc structure suggests the presence of more than one planet. For the central cavity alone (noted D1 in Figure~\ref{fig:sketch}), at least one planet is required to reproduce the cavity size \citep{Pohl+2017,Bertrang+2018,Toci+2020}. Moreover, \citet{Pohl+2017} and \citet{Toci+2020} modeled the gap D2 as the result of one super Jupiter planet, while \citet{Bertrang+2018,Bertrang+2020} modelled it with two Jupiter-like planets \ans{and an eccentrinc ring B1}. Finally, \citet{Perez+2019} could reproduce the B2, B3, B4, D3, and D4 features just adding one mini-Neptune at the B3 location (see Figure~\ref{fig:sketch}).

Although planet--disc interaction explains the radial structures in the disc, with one or more planets it is difficult to explain the misalignment of the inner disc (B0)
and the azimuthal structures at the innermost ring (B1). External gas inflows are able to produce a strong misalignment adding angular momentum \citep{Dullemond+2019,Kuffmeier+2020}, but there is no evidence of such inflows in HD~169142. Even though planets have been proposed as triggers of  azimuthal asymmetries in the dust  \citep[e.g. see][]{vanderMarel+2013, Perez+2014, Pacheco-Vazquez+2016, Fuente+2017}, the dust structures present at the B1 ring remain challenging to explain (we discuss this point in more detail in Section \ref{sec:discussion}). \ans{More recently, \citet{Bertrang+2020} and \citet{Montesinos+2021} proposed that the clumpiness and its apparent azimuthal rotation observed in \citet{Bertrang+2018},} could be due to a shadow caused by a circumplanetary disc and thermal instabilities. \ans{In addition, \citet{Toci+2020} found that the inner planet could be close to the inner edge of the B1 ring due to the dynamical interaction between the inner and the outer planet, and could be interacting with the dusty disc. 
}
Nevertheless those scenarios currently do not yet provide a compelling match to the observations.

Inspired by our previous work on dusty clumps in circumbinary discs \citep{Poblete+2019}, in this paper we show that the structures at the innermost ring B1 and the cavity size D1 of HD~169142 can be explained by the presence of a polar, moderately eccentric ($e_{\rm B}=0.2$), low mass-ratio ($q=0.1$) inner stellar-mass companion. \ans{The parameters chosen are in agreement with the current observational constrains for the system. The detectability of such a companion is discussed in further detail in Section \ref{sec:detectability}.} This binary hypothesis would change the nature of HD~169142 from a transitional disc to a circumbinary disc such as the already confirmed disc HD~142527 \citep{Price+2018} and potential binary systems such as IRS~48 \citep{Calcino+2019}, AB~Aurigae \citep{Poblete+2020}, and HD~143006 \citep{Ballabio+2021}. We describe the numerical methods and initial conditions in Section~\ref{sec:methods}. We report our results in Section~\ref{sec:results}. We discuss in Section~\ref{sec:discussion} and draw conclusions in Section~\ref{sec:conclusion}.


\section{Methods}
\label{sec:methods}

We perform 3D hydrodynamical simulations of circumbinary discs (CBDs) using the {\sc Phantom} smoothed particle hydrodynamics (SPH) code \citep{PricePH+2018b}. We use the dust-as-particles method in order to model the interaction between gas and dust particles as described in \citet{Laibe12a,Laibe12b} and \citet{Price&Laibe+2020}, which include the back-reaction effect of dust on gas. Our model consists of a central stellar binary, a circumbinary (P-type) planet and a circumbinary disc.

We model the two stars and the planet as sink particles \citep{Bate+1995}. The accretion radii are set to 1 au for the primary star, 0.5 au for the secondary star, and 1 Hill radius (3.5 au) for the planet. This choice of parameters avoids the formation of a circumplanetary and circumstellar discs, which are numerically expensive to simulate.

The binary and disc parameters are similar to those in \citet{Poblete+2019} for the polar alignment; the case that exhibits clumps very similar to HD~169142. We explored a limited number of combinations of semi-major axis, and eccentricity of the binary guided by the work previously mentioned to get the best values that reproduce the cavity size like the 1.3 mm observation \citep{Perez+2019}\footnote{We remark that the problem is degenerate and other sets of values of semi-major axis, mass, and/or eccentricity of the binary can give the same cavity size.}. The binary is initialised with a mass ratio of $q=M_2/M_1=0.1$ with $M_1=1.7\,M_{\odot}$ and $M_2=0.17\,M_{\odot}$. The companion is placed on an orbit with a semi-major axis of $11$~au, and an eccentricity $e_{\rm B}=0.2$, giving a period of roughly 26.68 yrs. The circumbinary disc is inclined by $i=90\degree$ with respect to the binary plane at the start of the simulation. The initial argument of periapsis ($\omega$) is set to $0\degree$ and $90\degree$ in order to cover the extreme orientations of the polar binary. As usual, the line of nodes ($\Omega$) is defined from the direction of the binary pericentre. It is set to $0\degree$ and $90\degree$ for the first and second case respectively. The planet has a mass of 2 $M_{\rm J}$ and is initialised on a circular and co-planar orbit with respect to the disc at 50 au, i.e. the planet is on a circumbinary orbit (P-type configuration).

The gas disc in our simulations is initialised with $5\times 10^5$ SPH particles in Keplerian rotation with a total gas mass of $0.01\ M_{\odot}$. It initially extends from $R_{\rm in}=22$ au to $R_{\rm out}=125$ au, with a power law surface density profile, $\Sigma\ \propto\ R^{-1}$. The disc is vertically isothermal and the temperature profile follows a shallower power law, $T\ \propto\ R^{-0.25}$, giving a scale height of $H/R = 0.048$ at $R_{\rm in}$ 
and $H/R = 0.074$ at $R_{\rm out}$. We set the SPH viscosity parameter $\alpha_{\rm AV} = 0.3$, which gives a mean \citet{Shakura&Sunyaev73} disc viscosity of $\alpha_{\rm SS}\approx5\times10^{-3}$ \citep[cf.][]{Lodato&Price2010}. The initial dust distribution coincides with the gas spatial distribution, but with a lower particle number equal to $1\times 10^5$. We consider a \hi{dust-to-gas ratio of 0.1, a grain size of 1 cm}, and an intrinsic grain density of $3\ \rm g\ cm^{-3}$. These parameters give an initial Stokes number of $\sim 2$ for the dust at the inner rim.

\ans{Due to the complexity of determining the inner rim size given the binary parameters \citep[see][]{Hirsh+2020}, our simulated ring size was $\sim$10\% larger than the observed one. Consequently, we set our code units to certain values to bring the simulation to a physical size that matches the observations: we reduced the length unit by 10\% and time units by $\sim$15\% in order to keep the gravitational constant invariant. Doing so, we now have a binary semi-major axis of 9.9 au with a period of 22.7 years and a planet with 45 au of semi-major axis. Note that the Stokes number is not affected by this change despite being 9 mm after the re-scaling. Hereafter we will refer to these re-scaled values and not to the original ones.}


\section{Results}
\label{sec:results}

The left and middle columns of Figure \ref{fig:simulation} show the disc morphology after approximately 230 binary and 45 planet orbits. The disc exhibits prominent gas spirals and dusty clumps. These asymmetric structures in the disc change periodically with the planet-binary location. These aspects will be discussed in the next sections. To aid with our discussion, we will use the regions defined in Figure \ref{fig:sketch}. \ans{Due to the large accretion radius used for the planet, at the end, the planet reaches a mass of $\sim3.85~M_{\rm J}$ and a semi-major axis of 43.46 au. } 

\subsection{Gas spirals}
\label{sec:gas}

The simulations exhibit multiple spiral arms in the gas distribution that can be classified spatially into three groups: those triggered by the planet, the ones located in the B1 ring, and those at the inner cavity D1.

The planet triggers two prominent spiral arms, one internal and one external to its orbit. Additional spiral arms with lower amplitude are also generated \citep{Dong+2015,Bae&Zhu2018a,Bae&Zhu2018b}, although only one additional outer spiral arm is visible in our simulations. The spiral arms interior to the planet interact with the spirals triggered by the stellar binary. The regions where the planet's spiral constructively interferes with the spirals originating from the internal binary result in higher gas densities. 

The binary-triggered spiral arms are located at the B1 and D1 regions. The simulations show that these spirals are densest close to the B1 ring. Our previous work shows an azimuthal concentration of those spiral arms near the apocenter of the binary \citep{Poblete+2019}. However, in this case, the inner planetary spiral concentrates most of gas in corotation with the planet. At the B1 ring, this corotation creates a sector with a gradient of angular velocity between the inner planetary spiral (sub-Keplerian movement) and the ones triggered by the binary, which can affect the dusty clumps. In addition, the planet prevents binary spirals to propagate beyond the D2-B1 region. These effects combined explain why there are no multiple spiral arms azimuthally concentrated as in HD~142527 \citep{Avenhaus+2014} and AB~Aurigae \citep{Fukagawa+2004,Hashimoto+2011}. 

\begin{figure*}%
\begin{center}
    
    \includegraphics[width=1.0\textwidth]{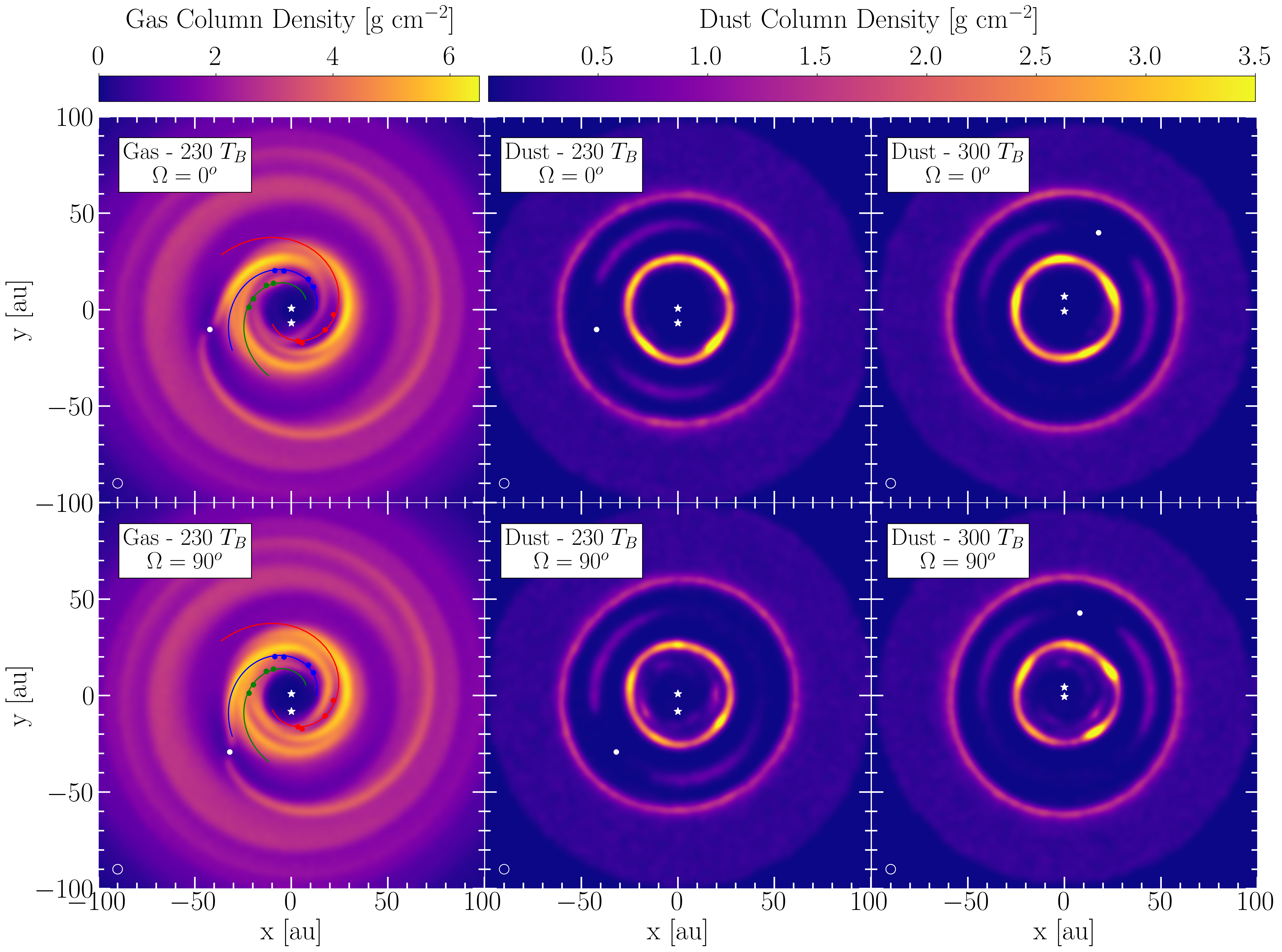} 
    \caption{
    Morphology for the simulation after 230 binary orbits and roughly 45 planet orbits for $\Omega=0\degree$ in the upper row and for $\Omega=90\degree$ in the lower row. The upper-left and lower-left panels show the gas disc, while the rest of panels show the dust disc. Observed spirals arms in HD~169142 (see Table \ref{tab:spiral_values}) are plotted along the gas morphology. The red spiral is the primary, the blue one is the secondary, and the green one is the tertiary. The respective colored points are those reported in Table \ref{tab:spiral_values}. The upper-right and lower-right panels show the dust morphology after 300 binary orbits for $\Omega=0\degree$ and $\Omega=90\degree$ respectively. The disc is seen face-on while the binary, located inside the cavity, is perpendicular to the disc. The white oval at the lower-left corner in all panels represent the smoothing size employed.
   } 
    \label{fig:simulation}
\end{center}
\end{figure*}

At the central cavity D1, we observe the third kind of spirals, often called streamers in binary models, which feed the central regions \citep{Dunhill+2015,Ragusa+2017,Price+2018,Poblete+2019,Hirsh+2020,Ragusa+2020,Ragusa+2021}. These spirals exhibit a correlation between their shape and the binary orbital phase. This property is useful for inferring a companion and its location \citep[see][for more details]{Poblete+2020}. Inside the cavity of HD~169142, \citet{Gratton+2019} report three spiral arms seen in NIR scattered light imaging, and they are summarized in Table \ref{tab:spiral_values}. The upper-left and lower-left panels of Figure \ref{fig:simulation} show the spatial distribution of the observed spiral superimposed on our simulations, which have been rotated in order to get a close match with the observed spirals. The primary arm (shown in red) starts at the lower part of the cavity while the secondary and the tertiary (shown in blue and green, respectively) start near the upper part. These spirals arms are expected to connect with a planet but the location of this putative planet is uncertain. We find a qualitatively good match with the observed position of the spirals. It is worth noting that, since this is a highly degenerate problem, a variety of matches can be found for different combinations; from different combinations between rotations and binary orbital phase to different combinations of the binary parameters such as semi-major axis, eccentricity, among others. The snapshots of our simulation are thus only one of many examples that match the observed morphology. One potential way to break this degeneracy is to compare the observed gas kinematics with those in our simulations. However, current ALMA data has not enough resolution to draw any conclusions along these lines \citep{Fedele+2017}. \ANS{Even the dedicated kinematic study by \citet{Yu+2021} only provides information for radii greater than 50 au, where the influence of the binary is negligible.}

\begin{table}
\centering
\begin{tabular}{cccccc}
\hline
\hline
Name & PA & R & pitch angle & $A$ & $\theta_o$\\
 & [degrees] & [mas] & [degrees] & [mas] & [degrees]\\
 
\hline
\multirow{ 4}{*}{Primary} 
    &196.7 & 157 & \multirow{ 4}{*}{15.3} & \multirow{ 4}{*}{36.3} & \multirow{ 4}{*}{69.2} \\
    &203.0 & 172\\
    &244.2 & 194\\
    &268.3 & 209\\

\hline
\multirow{ 4}{*}{Secondary} 
    &321.0 & 157 & \multirow{ 4}{*}{16.3} & \multirow{ 4}{*}{71.7} & \multirow{ 4}{*}{138.8}\\
    &335.9 & 172\\
    &16.0 & 194\\
    &28.0 & 209\\
    
\hline
\multirow{ 4}{*}{Tertiary} 
    &38.9 & 157 & \multirow{ 4}{*}{20.8} & \multirow{ 4}{*}{48.6} & \multirow{ 4}{*}{91.3}\\
    &50.9 & 172\\
    &79.0 & 194\\
    &91.6 & 209\\
    
\hline
\end{tabular}
\caption{Spiral position calculated in \citet{Gratton+2019} (see their table 6) and fit values. We consider a logarithmic spiral of the form $r(\theta)=Ae^{k(\theta-\theta_o)}$, where $k$ is the pitch angle, and $\theta$ the angular polar coordinate. We keep their computed pitch angle, and we just fit the rest of the logarithmic spiral parameters.}\label{tab:spiral_values}
\end{table}

\subsection{Dust structures}
\label{sec:dust}

Radially, the simulations exhibit two well defined dust rings \ANS{(see Figure \ref{fig:simulation})}, and the gaps D1 and D2. Note that we only produce one ring at roughly 67 au where rings B2, B3 and B4 are observed. Additionally, we observe dust in D2 at the coorbital region of the planet. The latter is densest closer to the planet and can be interpreted as dust trapped at the Lagrangian points $L_4$ and $L_5$ \citep{Montesinos+2020}. In order to reproduce the triple ring B2, B3, and B4 an additional migrating planet would be needed as proposed by \citet{Perez+2019}, where a 10 $M_{\earth}$ planet is able to split a single ring into three narrow rings.

The most outstanding aspect of the dust morphology is the formation of dusty clumps evenly distributed azimuthally at the B1 ring; similar to those reported in \citet{Poblete+2019}. Clump-like structures are observed in HD~169142 in multiple mm-continuum observations and our simulations display a dust distribution in B1 that is very similar to the observed one. Both simulations with $\Omega=0\degree$ and $\Omega=90\degree$ match reasonably well with the clumps' azimuthal separation. 


The clumps do not necessarily have a stable behaviour, see e.g. section 4.1 in \citet{Poblete+2019}. They continuously change in amplitude and position by the action of the gaseous spiral arms and their own rotation, respectively. The timescale of those changes is about one binary orbit. This would explain why the ALMA observations show only three clumps instead of four. \ans{Since one clump is needed to form a system of four clumps evenly distributed}; we can use the missed clump as an observational test in future observations (see Section \ref{sec:tests}). 

Finally, the case with $\Omega=90\degree$ exhibits dust concentrations just interior to the ring B1. These sub-structures are triggered by the gas spirals arms present in the cavity. These dust concentrations are significantly weaker in the case with $\Omega=0\degree$. 

\subsection{Density radial profiles}
\label{sec:cavity}

\begin{figure*}
\centering
\begin{center}
    {
    \begin{tabular}{c|c}
        \includegraphics[width=0.45\textwidth]{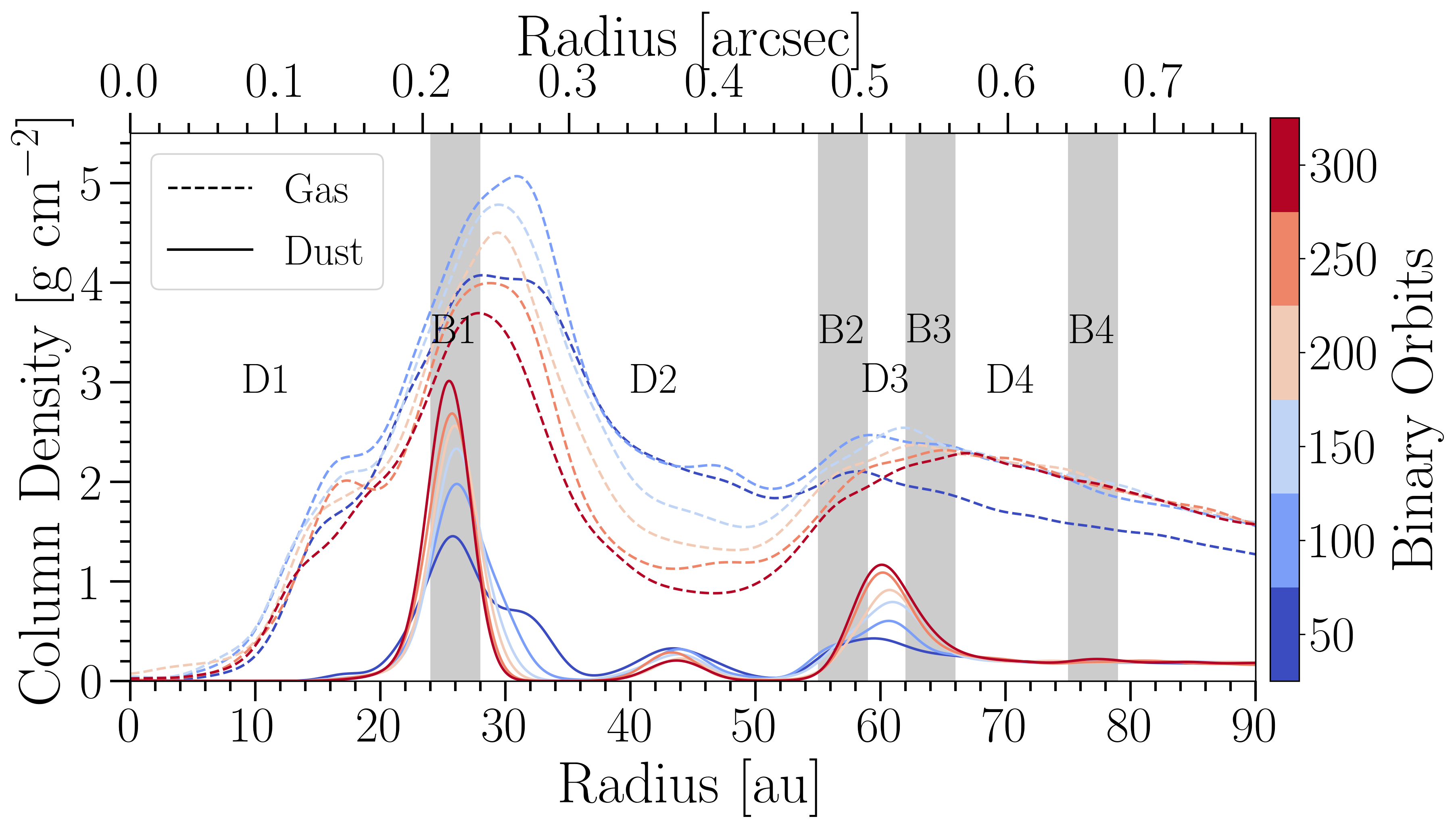} &  \includegraphics[width=0.45\textwidth]{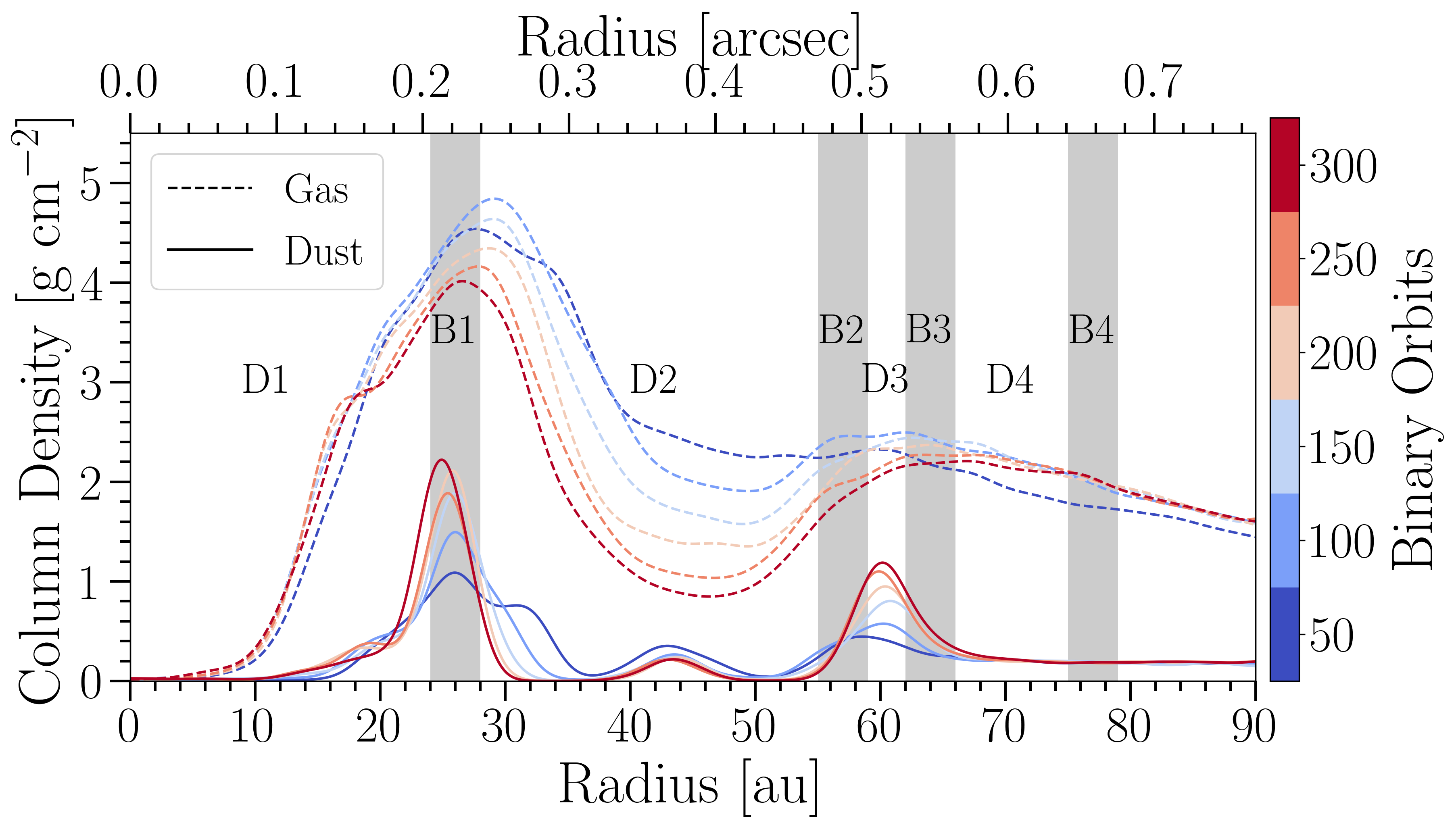}\\
        \includegraphics[width=0.45\textwidth]{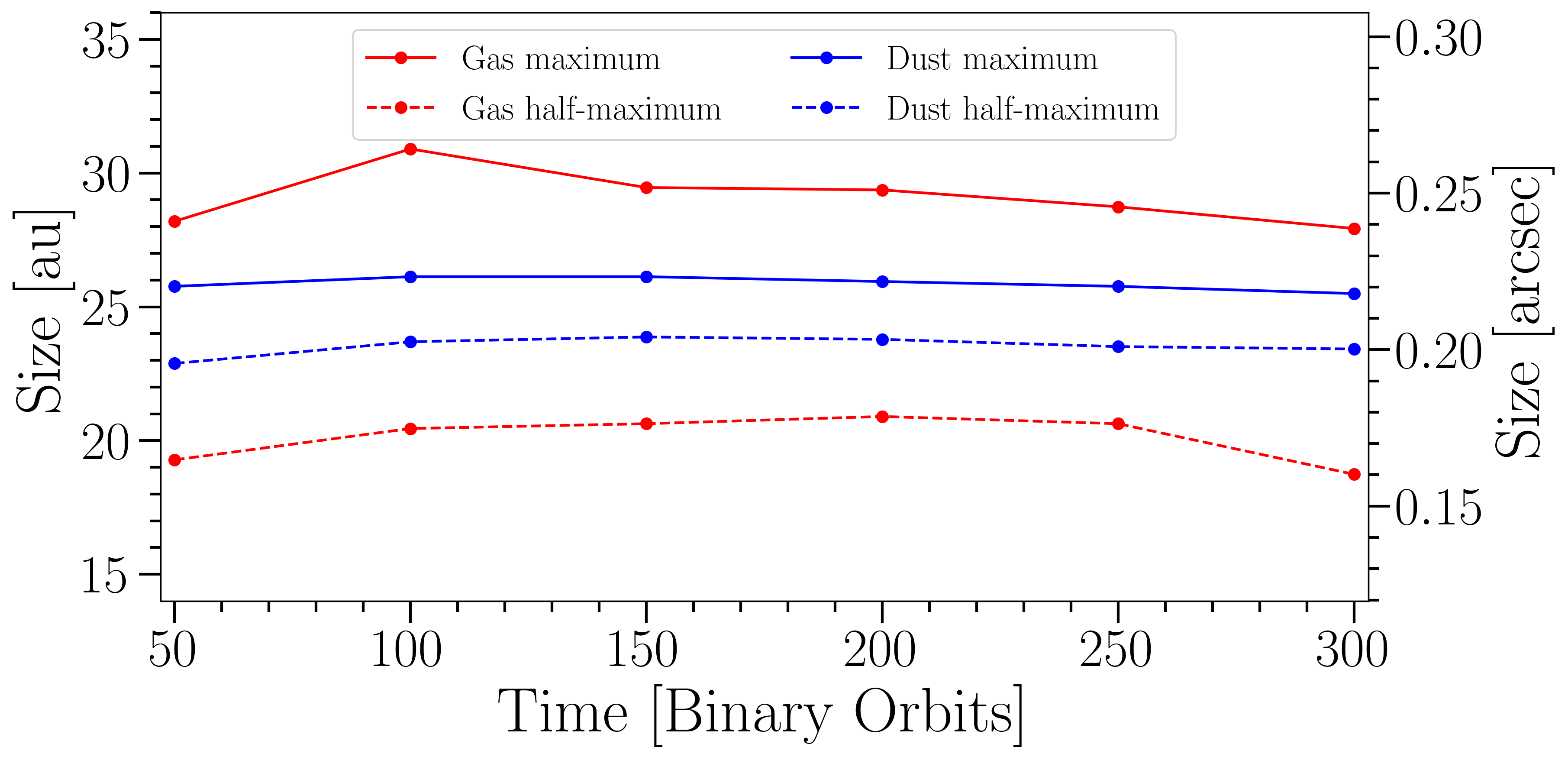} & \includegraphics[width=0.45\textwidth]{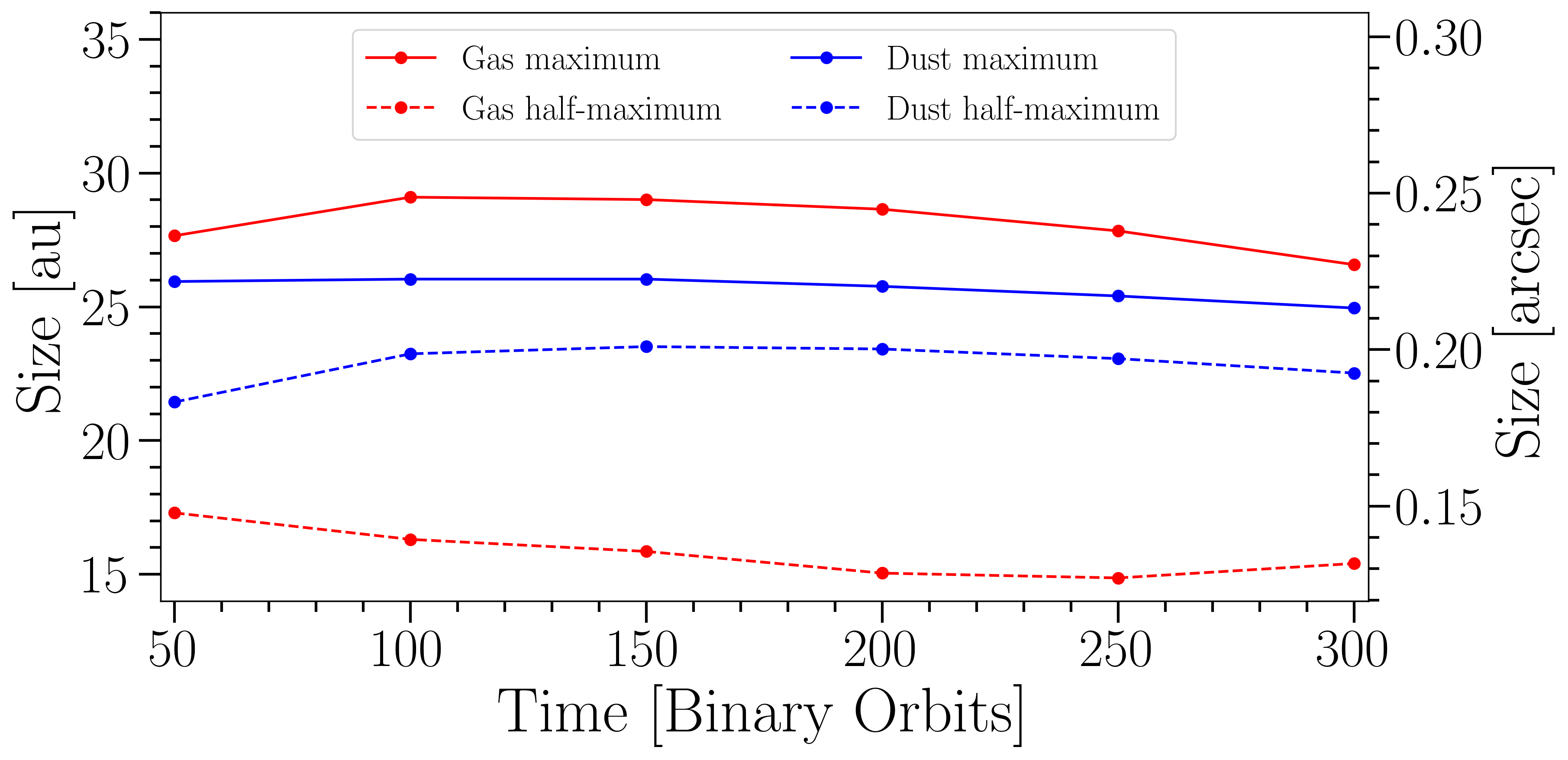}
    \end{tabular}
    \\
    }
    \caption{The upper panels show the evolution of the azimuthally-averaged radial profile of gas and dust for the case with $\Omega=0\degree$ (left) and with $\Omega=90\degree$ (right). The lower panels show the evolution of the cavity size of both dust and gas distribution, defined using different criteria. Red and blue lines represent the cavity size of gas and dust respectively. The cavity size is computed following, on one hand, the radius of the maximum of the density radial profile (solid line), and on the other hand, the radius at the half-maximum (dashed line).}
    \label{fig:cavity_size}
\end{center}
\end{figure*}

The azimuthally averaged radial profile and the cavity size for both gas and dust are displayed in Figure \ref{fig:cavity_size} for the case with $\Omega=0\degree$ and $\Omega=90\degree$. \ans{Note we are over-estimating the surface density by a factor of 10 in the B1 ring according to \citet{Fedele+2017}}. The gas surface density decreases with time at all radii, contrary to the dust surface density which increases in the ring B1 and the outer edge of the gap D2. Given the high Stokes number for the dust, it concentrates in the rings that correspond to gas pressure maxima \citep{Pinilla+2012a,Pinilla+2012b}. Changes in the inner cavity (D1) size are small, variations of a couple of au are observed in both gas and dust, with the strongest change in the gas between the first 50 and 100  binary orbits. The dust cavity size increases at the same period, then decreasing at a slow rate. Recent simulations have shown that cavity sizes can increase drastically in the long term; nevertheless, in polar configurations, the variations should remain at the level of a few au for our model parameters \citep{Hirsh+2020}. 

Finally, our simulations do not produce the circumstellar disc B0. This is likely because the large accretion radius and/or the high numerical viscosity inside the cavity triggers a rapid draining of the disc into the star given the large sink particle accretion radius chosen. Nevertheless, a circumstellar disc can be misaligned by a polar binary \citep[see figure 1 of][]{Poblete+2020}. Therefore, the binary scenario could also explain the misalignment of the inner disc B0. \ans{A planet in an inclined orbit could warp the inner rim of a disc \citep{Nealon+2018}. But in the B0 case, the proposed planet should be inside of the B0 disc and be massive enough to keep the disc inclined.} 

\subsection{Synthetic 1.3 mm observation}
\label{sec:synth}

\ans{In this Section we aim to compare the results from the SPH calculations with the ALMA observations by post-processing  our simulations using radiative transfer calculations. While the simulated grain size of 9 mm (see Section \ref{sec:methods}) is larger than the typical grain sizes traced by the ALMA observations ($\sim$1 mm), we note that the gas surface density in our simulations is 10 times higher than the values measured by \citet{Fedele+2017}, as we mentioned in the previous Section. Therefore, the behaviour exhibited by the 9 mm grains in our simulations will be equivalent to the one done by 0.9 mm grains since the Stokes number would be the same. We thus use the distribution of 9 mm grains in our simulations as tracer of 0.9 mm grains. }

\ans{We produce synthetic images using the Monte Carlo radiative transfer code \textsc{mcfost} \citep{pinte2006, pinte2009}. The
dust grain size distribution is set to a power-law with $dn/ds \propto s^{-3.5}$ for $0.03\ \mu$\textrm{m} $\leq s \leq 1$ mm, and a gas to dust ratio of 100. The gas mass is adopted from the simulations and we assume spherical and homogeneous dust grains composed of astronomical silicate with an intrinsic grain density of $3\ \rm g\ cm^{-3}$ \citep{weingartner2001, Li&Draine2001}. We assume the central star has an effective temperature $T_\textrm{eff} = 7650$ K and stellar radius of $R_\star = 1.65$ R$_\odot$ \citep{Meeus2010}, giving a blackblody luminosity of $\sim 8.4$ L$_\odot$, consistent with estimated values \citep{Dent2006}. The companion is assumed to have $T_\textrm{eff} = 3200$ K, $R_\star = 1.07$ R$_\odot$, giving a luminosity of $\sim 0.11$ L$_\odot$. Using $10^8$ Monte Carlo photon packets we computed the temperature and specific intensities at each wavelength, and produced images by ray-tracing the computed source function. } 


\ans{Since we include only a single dust grain size in our SPH simulations, assumptions are needed to complete the rest of the size spectrum.  We developed two different models, to cover a range of possible behaviours: 
\begin{itemize}
    \item RT model A: we assume that grains with sizes between $0.03$ $\mu$m and $100$ $\mu$m follow the gas particles, while grains between $100$ $\mu$m and 1 mm follow the dust particles. 
    \item RT model B: we assume that grains with sizes between $0.03$ $\mu$m and $500$ $\mu$m follow the gas particles, while grains between $500~\mu$m and 1 mm follow the dust particles.
\end{itemize}
These radiative transfer models are meant to illustrate possible scenarios for the dust distribution in HD169142.  The scalability of our model allows for this qualitative exploration. We note that we are not claiming that we can constrain the grain sizes with these models. } 


\ans{Figure \ref{fig:RT} displays two different radiative transfer calculations for both simulations after 230 binary orbits together with the 1.3~mm ALMA observation. The observed ring morphology is better reproduced in $\Omega=0\degree$ than in $\Omega=90\degree$ case. But the case $\Omega=90\degree$ can trap dust just interior to the ring B1. If we compare it with the 1.3mm ALMA observation, an intermediate scenario could be the solution.} 

\ans{If we just concentrate in the case $\Omega=0\degree$, the RT model B exhibits dust that is coupled to the gas highlighting the gas structures too prominently (disc and inner spiral arms, see Section \ref{sec:gas}). On the other hand, such features are completely suppressed in RT~A as the emission is dominated by the large grains that are mostly concentrated in the ring B1. Taken individually, both our radiative models fail to exactly reproduce the fainter emission just interior to the ring seen in the ALMA observation. That fainter emission, however, has a morphology that is similar to the dust distribution seen close to the inner rim in RT B. Thus, we speculate that it could be produced by grains smaller than $\approx1$~mm, slightly coupled to the gas, and still emitting significantly at 1.3~mm.} 

\ans{All in all, reproducing the ALMA observation would require a multi-dust species simulation, which is beyond the scope of this work.} 
\section{Discussion}
\label{sec:discussion}

Previous studies of HD~169142 had only focused on planet--disc interactions and had neglected the possibility of a stellar companion in the central cavity. Even though one or more planets are probably needed to reproduce the outer rings and gaps, dust structures in the innermost ring suggest a companion in the stellar mass regime. As shown in Section \ref{sec:results}, the polar stellar binary has the capability of simultaneously triggering the formation of a large cavity, dusty clumps, and spirals.

\subsection{Nature of the companion}
\label{sec:companion_nature}

\subsubsection{Effects of the mass companion}

Binary--planet interactions are able to explain the complex morphology of some protoplanetary discs. Recently \citet{Ballabio+2021} proposed an inclined binary with a P-type planet to explain reasonably well the HD~143006 disc structures. \ans{Additionally, a binary-planet system is stable in long-term evolution \citep{Ballantyne+2021}.} Although we have shown that the polar binary hypothesis could reproduce the non-axisymmetric features seen in the HD~169142's disc, there are a few important considerations to evaluate the likelihood of such scenario. It seems reasonable to assume that one or more planets are inside the central cavity D1 due to the system's age and the lack of evidence for a stellar companion. Nevertheless, in this particular case, planets have not been shown to reproduce the observed features thus far.

First of all, a typical planetary candidate would be co-planar and with an almost circular orbit. Such a planet configuration has already been considered in order to reproduce the gas disc of HD~169142 \citep{Pohl+2017,Bertrang+2018}, giving a good match for the density profile. To our knowledge, only \citet{Toci+2020} have modelled the B1 dust ring with a similar Stokes to our work, and their simulations result in a central cavity like the observed one, but without any substructure or clumping at the innermost dust ring\footnote{Note that \citet{Toci+2020} use {\sc Phantom} as well, which is not able to produce vortices given the high viscosity. Including such vortices, the B1 ring could exhibit sub-structures.}.


If we consider a companion with a mass ratio $q=0.01$, an order of magnitude lower than what we propose in this work and coplanar, the dust ring B1 should be quite axisymmetric \citep[e.g. see figure 2 of][]{Ragusa+2017}, provided the disc viscosity is high enough to subdue the Rossby-wave instability. Additionally, if the mass ratio is increased further then an asymmetric ``horseshoe'' feature begins to develop around the cavity \citep{Ragusa+2017, Ragusa+2020}. Such a structure is not observed in HD~169142 \citep{Perez+2019}. Another problem that arises with increasing the mass ratio of a coplanar companion is that the cavity begins to become eccentric \citep{Ragusa+2020}, leading to the central source being misaligned with the projected centre of the cavity. But the central source in HD~169142 is well-centered in the cavity \citep{Perez+2019}. These facts suggest either a low-mass inner companion, or a companion misaligned with respect to the outer disc. Indeed, a polar orientation can produce an alignment between both bodies in the sky-plane, appearing to us as a  well-centered system for low disc inclinations. 

We highlight that the scenario where there are two or more planets inside the cavity has not been explored. Planets are able to induce dust trapping in a vortex \citep{Ataiee+2013, Birnstiel+2013, Lyra&Lin2013}, therefore there could be a special configuration that reproduces more than one dust-trap in the same ring. Nevertheless, alternative models should consider the particular azimuthal separation of the observed clumps in HD~169142.

\subsubsection{Detectability}
\label{sec:detectability}

\begin{figure*}
\centering
\begin{center}

    \begin{tabular}{c|c}
        \includegraphics[width=0.6\textwidth]{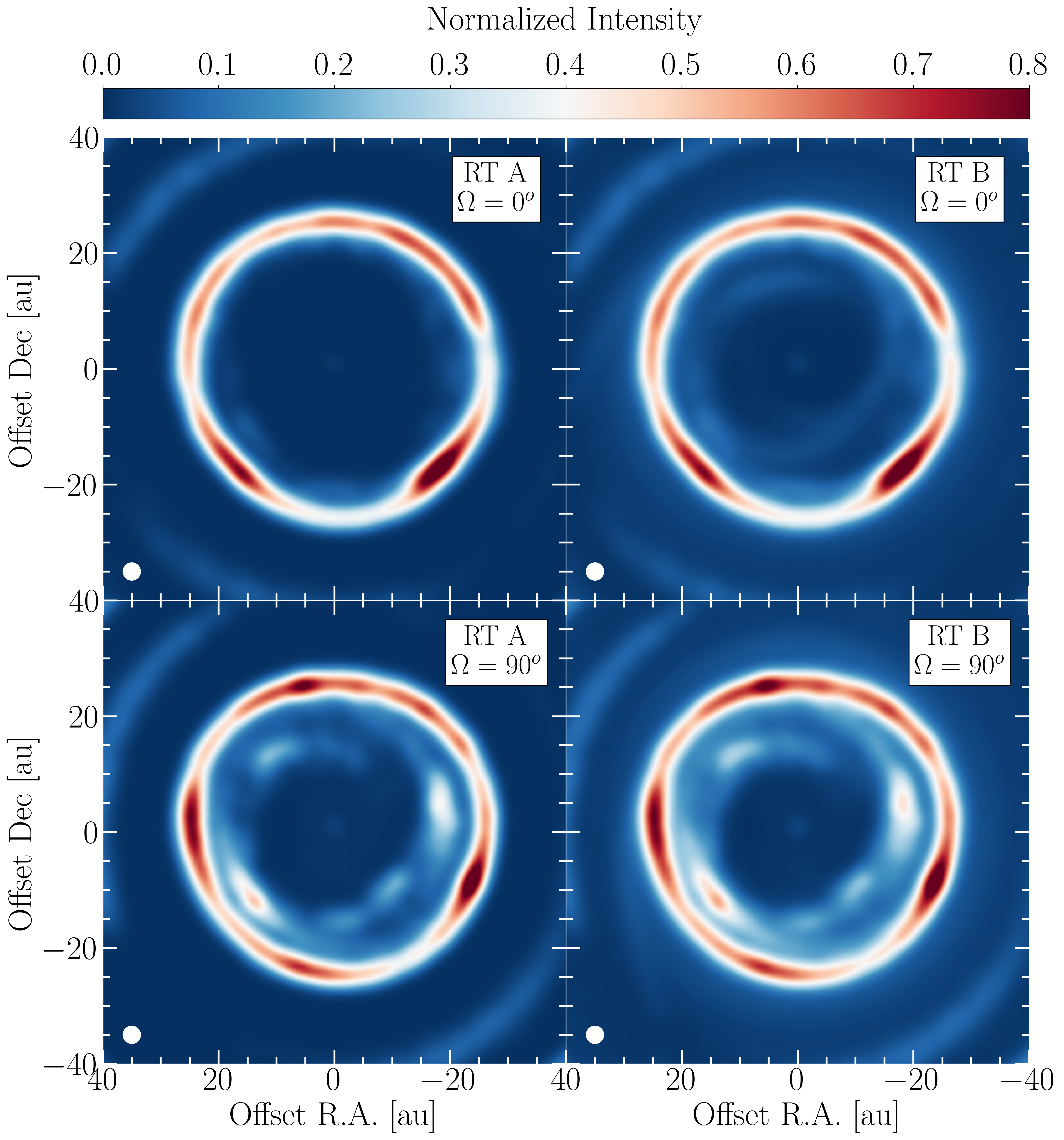} &  \includegraphics[width=0.34\textwidth]{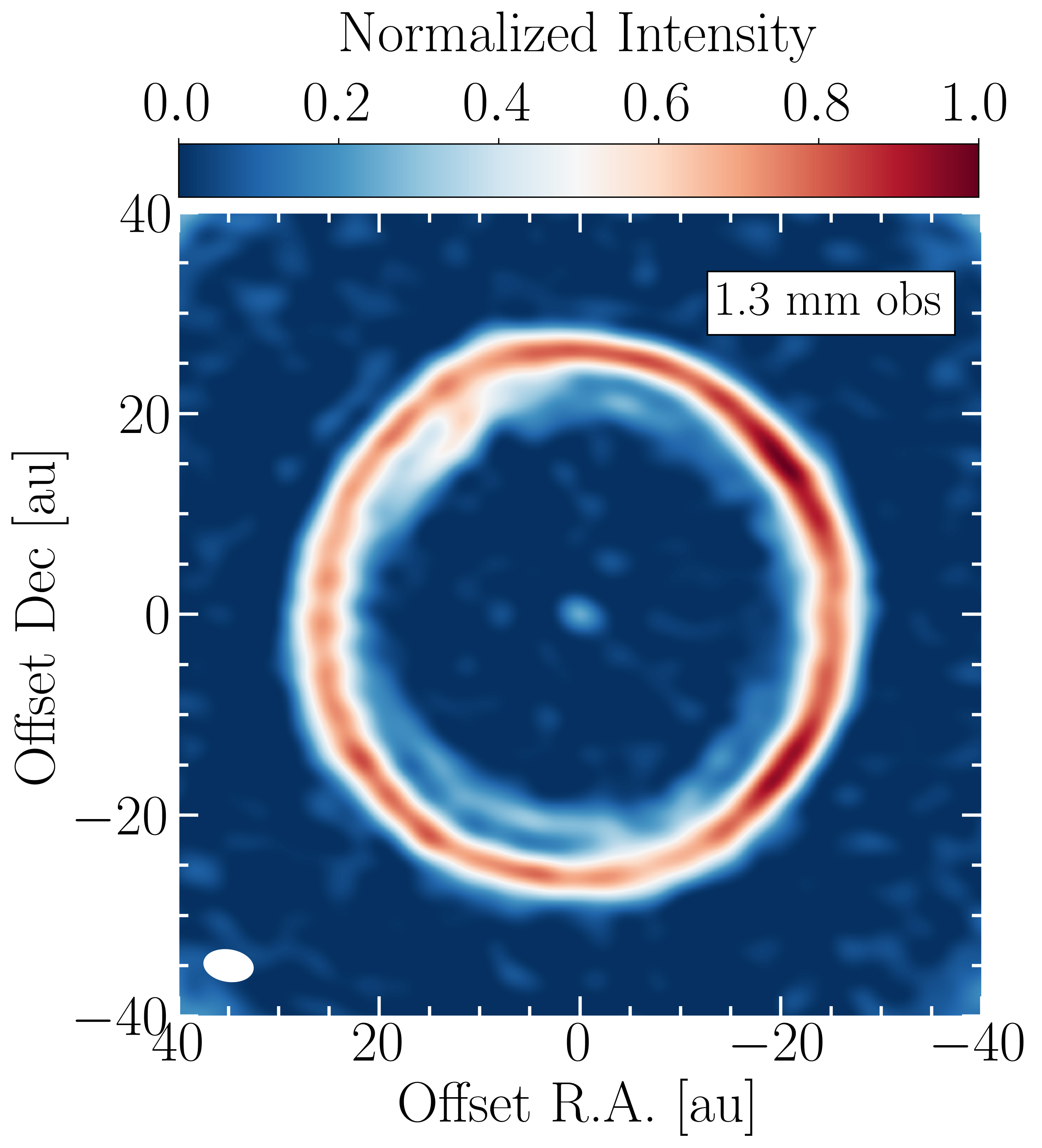}
    \end{tabular}

    \caption{\ans{\textbf{Left figure:} radiative transfer calculations for the simulations. The left column shows the RT model A while the right column shows the RT model B. The upper row shows the simulation for the case with $\Omega=0\degree$ while the lower row shows the simulation for the case with $\Omega=90\degree$. \textbf{Right figure:} 1.3 mm ALMA observation. The intensity is normalised to the peak value for each panel.}} 
    \label{fig:RT}
\end{center}
\end{figure*}

Overall, our proposed binary can reproduce reasonably well the cavity size, the clump structure, and the presence of gaseous spiral arms. However, the system has been studied for a long time, and no stellar companion within the central cavity has been observed so far \citep[see][for a summary of the observational constraints]{vanderMarel+2021}.

\begin{figure}
\centering
\begin{center}

    \includegraphics[width=0.45\textwidth]{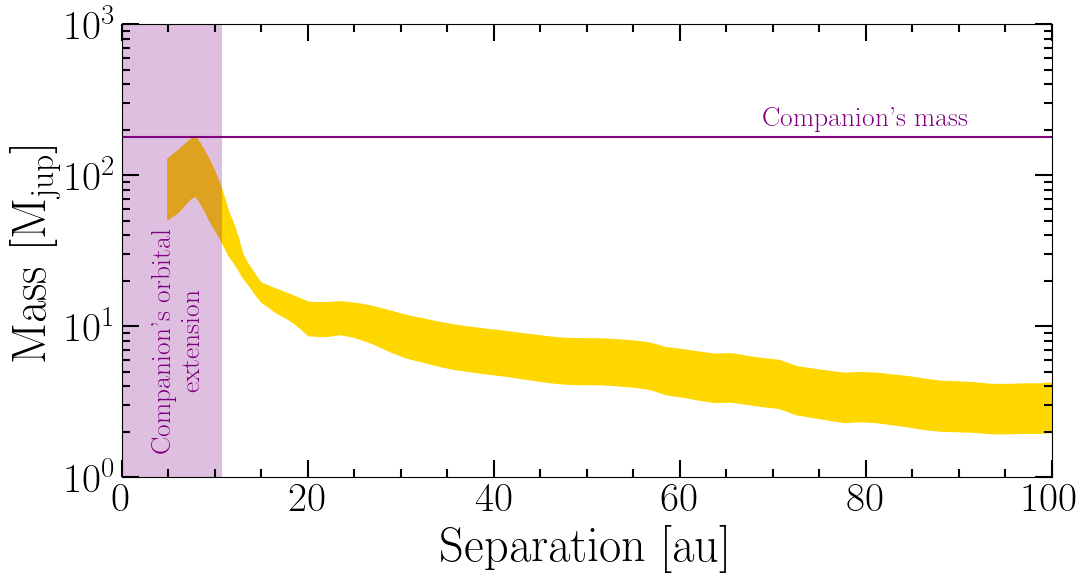} 
    \caption{ Contrast limits translated into upper limit on the mass from high-contrast imaging observations. The upper mass limit is displayed in yellow and the binary parameters modelled are displayed in purple.}
    \label{fig:observational_constraints}
\end{center}
\end{figure}

\citet{Ligi+2018} presented high-contrast imaging data for the source. From their contrast curves obtained with SPHERE/IFS in April 2017 (without coronagraph) we can derive mass upper limits for any putative companion beyond the inner working angle. To obtain the limits shown in Figure \ref{fig:observational_constraints}, we converted the contrast limits into mass using an age for the system of 6$^{+6}_{-3}$ Myr and the evolutionary models BT-SETTL \citep{Allard+2013}. 

Additionally, the target was observed by the $4.1$ m Southern Astrophysical Research Telescope (SOAR) with the speckle camera, HRCam \citep{Tokovinin2018} in March 2021. Two data cubes were taken, to exclude any sporadic phenomena. Assuming the same age and evolutionary model used in the SPHERE/IFS contrast curves, this speckle interferometry observation implies a non-detection limit for a companion $\gtrsim 400$ M$_{\rm J}$ at separations $>4.6$ au. 

The proposed inner companion in this work would be at the limit of the detection. Figure \ref{fig:observational_constraints} displays the best current constraints from direct imaging observations. Interior to 4.7 au, the inner working angle, the constraints are very poor and the proposed companion would have been missed. For both binary scenarios we explore, we find that there is a large ($\sim 35$\%) probability that the binary could have remained hidden and undetected at such small separations. Elsewhere along its orbital path, the companion would lie above the detection limit. Nevertheless, the proposed companion mass is very close to the limits exposed in Figure \ref{fig:observational_constraints} and thus its detection would be challenging. In addition, we find that there is a wide parameter space in mass, semi-major axis, and eccentricity in which the binary could roughly reproduce the morphology of the B1 ring. Future monitoring of this system could confirm our proposed scenario if the companion is detected, or provide additional multi-epoch constraints if it remains undetected.  

Finally, it is worth mentioning there are no strong spectroscopic constraints of this system, which are ideal to assess the presence of binaries at close separations ($\lesssim$10 au) like the one we are proposing. \ans{However, the spectral features associated to A and F types together with the unusual X-ray emission observed in HD~169142 lend support to the binary scenario \citep{Grady+2007}}. In summary, our proposed binary is consistent with the current published constraints for this system, and our scenario can be tested with further monitoring.

\subsection{Future observational tests}
\label{sec:tests}

The dust clumps can evolve over time, as shown by \citet{Poblete+2019}. This effect can be appreciated comparing the dust panels (middle and right columns) of Figure \ref{fig:simulation} where clumps can vanish at some epoch. We would be observing an epoch where only three clumps are at the B1 ring. As we discussed in Section \ref{sec:dust} a single clump can disappear in roughly one binary orbit (22.7 years) and form again in the same period. Since strong variations happen in timescales that are similar to the binary period, future observations covering a baseline of 20~years should be able to confirm our binary hypothesis. If the binary exists, a new clump can appear or disappear in the B1 ring and this can be tested by monitoring this system over the next 23 years.

\begin{figure}
\centering
\begin{center}

    \includegraphics[width=0.45\textwidth]{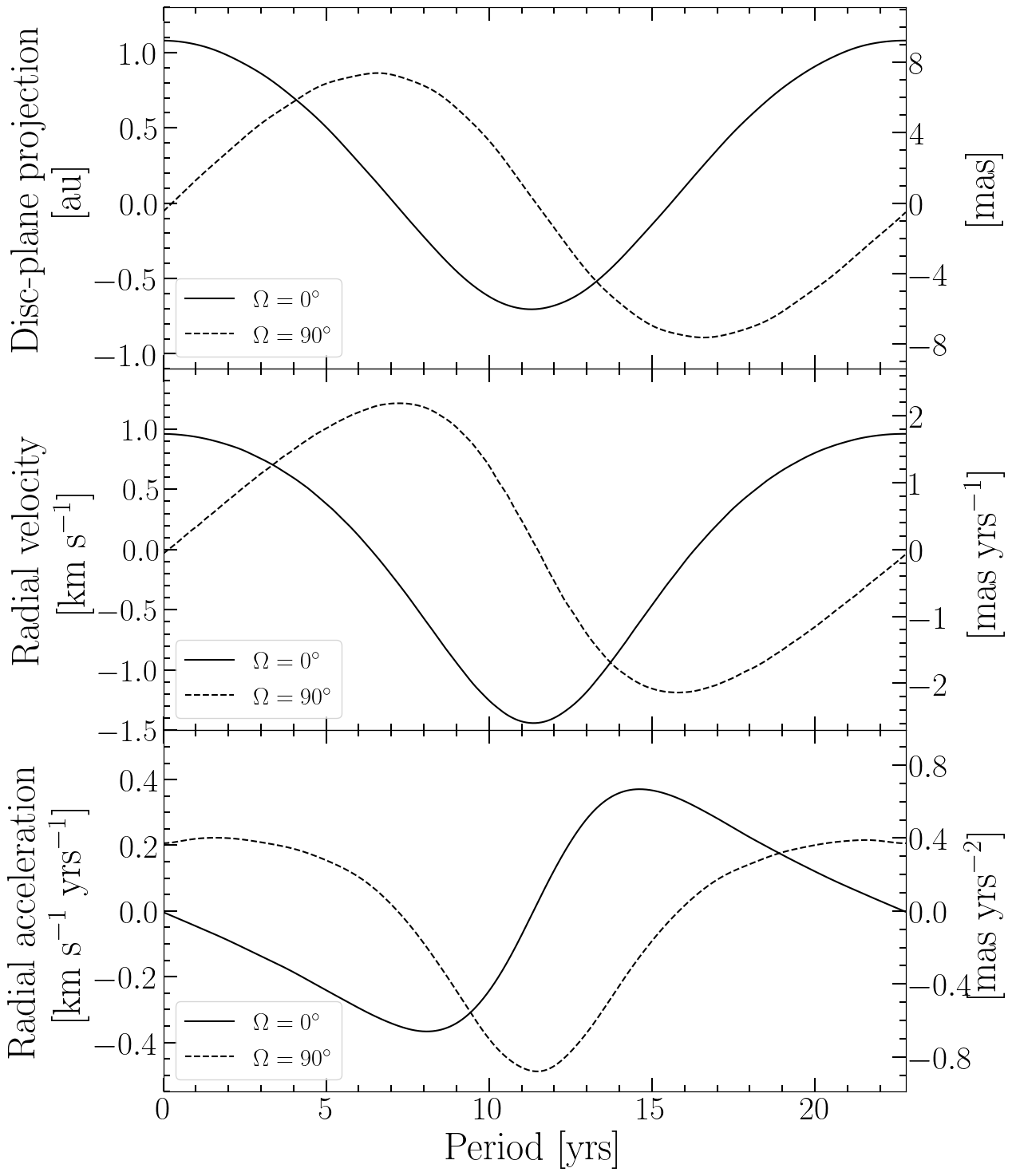} 
    \caption{Observable tests for the primary star. The upper panel shows the movement of the central star in the disc-plane, the middle panel shows the radial velocity, and the lower panel shows the radial acceleration for the primary star. We are assuming that the disc is completely face-on and the binary is perpendicular to it to build the panels. The case with $\Omega=0\degree$ is displayed as a solid line, while the case with $\Omega=90\degree$ is displayed as a dashed line in both panels.}
    \label{fig:astrometry}
\end{center}
\end{figure}

Also, given \textit{Gaia}'s astrometric precision of 20 $\mu$as \citep{Gaia2018} it should be possible to determine in the future if HD~169142 has an astrometric wobble that can be attributed to a companion more massive than a planet in the cavity. If the companion exists, the primary star HD~169142 should show a projected sky-plane proper motion similar to the upper panel of Figure \ref{fig:astrometry}. Indeed, \citet{Gaia+2021} give promising preliminary results for the astrometric excess noise that would favour our model: HD~169142 has 0.12 mas at $5.7\sigma$ level, similar to the 0.16 mas that HD~142527 has, a known $q\sim0.1$ binary. Furthermore,  long-term radial velocity measurements could confirm or rule out the binary scenario as well. \ans{It is worth noting that the presence of the circumstellar disc could give fake astrometric signals if there are shadows or other illumination effects.} The middle and lower panels of Figure \ref{fig:astrometry} display the changes in the radial velocity and the radial acceleration of the primary star, respectively. Monitoring over 5 years should be enough to detect a RV trend of $\sim$0.2 km s$^{-1}$ year$^{-1}$.

In reality, since HD~169142's disc is not perfectly face-on, the binary on a polar configuration will not be seen perfectly edge-on, so the effect on the radial velocity will be slightly lower than our estimation. Nevertheless, due to the low inclination between the disc of HD~169142 and the sky-plane, Figure \ref{fig:astrometry} is a reasonable approximation.


	
\section{Conclusion}
\label{sec:conclusion}

We performed 3D SPH gas and dust simulations of circumbinary discs to explain the properties of HD~169142. Our models show that an unseen stellar companion inside the disc cavity can reproduce many of the observed features in this system. If future observations detect such a companion, the disc around HD~169142 should be termed as \textit{circumbinary} instead of \textit{transitional}. Our main findings are summarised as follows:

\begin{enumerate}
    \item A stellar companion in a polar orbit is able to reproduce the dust structures (dust clumps and cavity size) observed in HD~169142. \ans{Nevertheless, in order to reproduce the faint dust emission just interior to the ring, a multi-grain simulation is needed.}
    
    \item The companion produces spiral arms that coincide in scale with the observed ones at the disc cavity. But, due to the degeneracy between disc rotation and evolutionary stage of the binary in one orbit, we cannot provide a stronger orbit constraint. In addition, one of the spirals could connect with the external planet location. 
    
    \item We expect future observations to show changes in the dust innermost ring morphology.
    
    \item According to our binary model, we expect future changes in the astrometric parameters of the star HD~169142 \ans{. Specifically, we predict radial velocity variations of $\sim$0.2 km s$^{-1}$ year$^{-1}$.} 
    
\end{enumerate}

Our work follows the re-interpretation of some protoplanetary discs, such as HD~142527 \citep{Price+2018}, IRS~48 \citep{Calcino+2019}, AB~Aurigae \citep{Poblete+2020}, and HD~143006 \citep{Ballabio+2021} where stellar-mass companions were proposed to explain the observed disc features. In this case, we are also considering a planet further out. Our results support the presence of a massive body and (at least) one P-type planet producing the dust structures in HD~169142. Future observations in the next two decades of the inner cavity and/or astrometric measurements could confirm or rule out our proposed polar binary.

            
\section*{Acknowledgements}

We thank A.~Tokovinin for promptly observing this system with SOAR, as reported in \S~\ref{sec:detectability}.
This work has been partially supported by the Deutsche Forschungsgemeinschaft (grant LO 1715/2-1, within Research Unit FOR 2285 ``Debris Disks in Planetary System''. 
This project has received funding from the European Union's Horizon 2020 research and innovation programme under the Marie Sk\l{}odowska-Curie grant agreements Nº 210021 and Nº 823823 (DUSTBUSTERS). 
JCa acknowledges support from an Australian Government Research Training Program Scholarship, and the Laboratory Directed Research and Development Program (LDRD) (approved for public release as LA-UR-21-24101). 
SP y AZ acknowledge support from Millennium Nucleus NCN2021\_080.
JCu, MM, SZF, and AB acknowledge support by ANID, -- Millennium Science Initiative Program -- NCN19\_171. 
S.Z-F acknowledges financial support from the European Southern Observatory via its studentship program and ANID via PFCHA/Doctorado Nacional/2018-21181044.
S.P. acknowledges support from ANID Fondecyt Regular grant 1191934. 
A.Z. acknowledges support from the FONDECYT Iniciaci\'on en investigaci\'on project number 11190837.
S.M. acknowledges support from a Junior Research Fellowship from Jesus College, Cambridge.
The Geryon2 cluster housed at the Centro de Astro-Ingenier\'ia UC was used for the calculations performed in this paper. The BASAL PFB-06 CATA, Anillo ACT-86, FONDEQUIP AIC-57, and QUIMAL 130008 provided funding for several improvements to the Geryon/Geryon2 cluster. CP and DJP acknowledge Australian Research Council funding via FT170100040 and DP180104235.

\section*{DATA AVAILABILITY}

The data underlying this article will be shared on reasonable request to the corresponding author. The code {\sc Phantom} used in this work is publicly available at \url{https://github.com/danieljprice/phantom}. {\sc mcfost} is available on request. The reduced ALMA image that we used in Figure~\ref{fig:RT} is directly available for download from \citet{Perez+2019}. The SPHERE data are publicly available and can be queried and downloaded directly from \url{http://archive.eso.org/}. The SOAR data were provided by Andrei Tokovinin under his permission. Such data will be shared on request to the corresponding author with the permission of Andrei. The data will be available without requests in early 2022 in \url{http://www.ctio.noao.edu/~atokovin/papers/}.

\bibliographystyle{mnras}
\bibliography{HD169142}

\begin{thebibliography}{}
\makeatletter
\relax
\def\mn@urlcharsother{\let\do\@makeother \do\$\do\&\do\#\do\^\do\_\do\%\do\~}
\def\mn@doi{\begingroup\mn@urlcharsother \@ifnextchar [ {\mn@doi@}
  {\mn@doi@[]}}
\def\mn@doi@[#1]#2{\def\@tempa{#1}\ifx\@tempa\@empty \href
  {http://dx.doi.org/#2} {doi:#2}\else \href {http://dx.doi.org/#2} {#1}\fi
  \endgroup}
\def\mn@eprint#1#2{\mn@eprint@#1:#2::\@nil}
\def\mn@eprint@arXiv#1{\href {http://arxiv.org/abs/#1} {{\tt arXiv:#1}}}
\def\mn@eprint@dblp#1{\href {http://dblp.uni-trier.de/rec/bibtex/#1.xml}
  {dblp:#1}}
\def\mn@eprint@#1:#2:#3:#4\@nil{\def\@tempa {#1}\def\@tempb {#2}\def\@tempc
  {#3}\ifx \@tempc \@empty \let \@tempc \@tempb \let \@tempb \@tempa \fi \ifx
  \@tempb \@empty \def\@tempb {arXiv}\fi \@ifundefined
  {mn@eprint@\@tempb}{\@tempb:\@tempc}{\expandafter \expandafter \csname
  mn@eprint@\@tempb\endcsname \expandafter{\@tempc}}}

\bibitem[\protect\citeauthoryear{{Alexander}, {Clarke}  \&
  {Pringle}}{{Alexander} et~al.}{2006}]{Alexander+2006}
{Alexander} R.~D.,  {Clarke} C.~J.,   {Pringle} J.~E.,  2006, \mn@doi [\mnras]
  {10.1111/j.1365-2966.2006.10293.x}, \href
  {https://ui.adsabs.harvard.edu/\#abs/2006MNRAS.369..216A} {369, 216}

\bibitem[\protect\citeauthoryear{{Alexander}, {Pascucci}, {Andrews}, {Armitage}
   \& {Cieza}}{{Alexander} et~al.}{2014}]{Alexander+2014}
{Alexander} R.,  {Pascucci} I.,  {Andrews} S.,  {Armitage} P.,   {Cieza} L.,
  2014, \mn@doi [Protostars and Planets VI]
  {10.2458/azu_uapress_9780816531240-ch021}, \href
  {http://adsabs.harvard.edu/abs/2014prpl.conf..475A} {pp 475--496}

\bibitem[\protect\citeauthoryear{{Allard}, {Homeier}  \& {Freytag}}{{Allard}
  et~al.}{2013}]{Allard+2013}
{Allard} F.,  {Homeier} D.,   {Freytag} B.,  2013, \memsai, \href
  {https://ui.adsabs.harvard.edu/abs/2013MmSAI..84.1053A} {84, 1053}

\bibitem[\protect\citeauthoryear{{Ataiee}, {Pinilla}, {Zsom}, {Dullemond},
  {Dominik}  \& {Ghanbari}}{{Ataiee} et~al.}{2013}]{Ataiee+2013}
{Ataiee} S.,  {Pinilla} P.,  {Zsom} A.,  {Dullemond} C.~P.,  {Dominik} C.,
  {Ghanbari} J.,  2013, \mn@doi [\aap] {10.1051/0004-6361/201321125}, \href
  {https://ui.adsabs.harvard.edu/abs/2013A&A...553L...3A} {553, L3}

\bibitem[\protect\citeauthoryear{{Avenhaus}, {Quanz}, {Schmid}, {Meyer},
  {Garufi}, {Wolf}  \& {Dominik}}{{Avenhaus} et~al.}{2014}]{Avenhaus+2014}
{Avenhaus} H.,  {Quanz} S.~P.,  {Schmid} H.~M.,  {Meyer} M.~R.,  {Garufi} A.,
  {Wolf} S.,   {Dominik} C.,  2014, \mn@doi [\apj]
  {10.1088/0004-637X/781/2/87}, \href
  {https://ui.adsabs.harvard.edu/#abs/2014ApJ...781...87A} {781, 87}

\bibitem[\protect\citeauthoryear{{Bae} \& {Zhu}}{{Bae} \&
  {Zhu}}{2018a}]{Bae&Zhu2018a}
{Bae} J.,  {Zhu} Z.,  2018a, \mn@doi [\apj] {10.3847/1538-4357/aabf8c}, \href
  {https://ui.adsabs.harvard.edu/abs/2018ApJ...859..118B} {859, 118}

\bibitem[\protect\citeauthoryear{{Bae} \& {Zhu}}{{Bae} \&
  {Zhu}}{2018b}]{Bae&Zhu2018b}
{Bae} J.,  {Zhu} Z.,  2018b, \mn@doi [\apj] {10.3847/1538-4357/aabf93}, \href
  {https://ui.adsabs.harvard.edu/abs/2018ApJ...859..119B} {859, 119}

\bibitem[\protect\citeauthoryear{{Bae}, {Zhu}  \& {Hartmann}}{{Bae}
  et~al.}{2017}]{Bae+2017}
{Bae} J.,  {Zhu} Z.,   {Hartmann} L.,  2017, \mn@doi [\apj]
  {10.3847/1538-4357/aa9705}, \href
  {https://ui.adsabs.harvard.edu/abs/2017ApJ...850..201B} {850, 201}

\bibitem[\protect\citeauthoryear{{Bailer-Jones}, {Rybizki}, {Fouesneau},
  {Mantelet}  \& {Andrae}}{{Bailer-Jones} et~al.}{2018}]{Bailer-Jones+2018}
{Bailer-Jones} C.~A.~L.,  {Rybizki} J.,  {Fouesneau} M.,  {Mantelet} G.,
  {Andrae} R.,  2018, \mn@doi [\aj] {10.3847/1538-3881/aacb21}, \href
  {https://ui.adsabs.harvard.edu/abs/2018AJ....156...58B} {156, 58}

\bibitem[\protect\citeauthoryear{{Ballabio}, {Nealon}, {Alexander}, {Cuello},
  {Pinte}  \& {Price}}{{Ballabio} et~al.}{2021}]{Ballabio+2021}
{Ballabio} G.,  {Nealon} R.,  {Alexander} R.~D.,  {Cuello} N.,  {Pinte} C.,
  {Price} D.~J.,  2021, \mn@doi [\mnras] {10.1093/mnras/stab922}, \href
  {https://ui.adsabs.harvard.edu/abs/2021MNRAS.tmp..913B} {}

\bibitem[\protect\citeauthoryear{{Ballantyne} et~al.,}{{Ballantyne}
  et~al.}{2021}]{Ballantyne+2021}
{Ballantyne} H.~A.,  et~al., 2021, \mn@doi [\mnras] {10.1093/mnras/stab2324},
  \href {https://ui.adsabs.harvard.edu/abs/2021MNRAS.507.4507B} {507, 4507}

\bibitem[\protect\citeauthoryear{{Bate}, {Bonnell}  \& {Price}}{{Bate}
  et~al.}{1995}]{Bate+1995}
{Bate} M.~R.,  {Bonnell} I.~A.,   {Price} N.~M.,  1995, \mn@doi [\mnras]
  {10.1093/mnras/277.2.362}, \href
  {https://ui.adsabs.harvard.edu/#abs/1995MNRAS.277..362B} {277, 362}

\bibitem[\protect\citeauthoryear{{Bertrang}, {Avenhaus}, {Casassus},
  {Montesinos}, {Kirchschlager}, {Perez}, {Cieza}  \& {Wolf}}{{Bertrang}
  et~al.}{2018}]{Bertrang+2018}
{Bertrang} G.~H.~M.,  {Avenhaus} H.,  {Casassus} S.,  {Montesinos} M.,
  {Kirchschlager} F.,  {Perez} S.,  {Cieza} L.,   {Wolf} S.,  2018, \mn@doi
  [\mnras] {10.1093/mnras/stx3052}, \href
  {https://ui.adsabs.harvard.edu/abs/2018MNRAS.474.5105B} {474, 5105}

\bibitem[\protect\citeauthoryear{{Bertrang}, {Flock}, {Keppler}, {Trifonov},
  {Penzlin}, {Avenhaus}, {Henning}  \& {Montesinos}}{{Bertrang}
  et~al.}{2020}]{Bertrang+2020}
{Bertrang} G. H.~M.,  {Flock} M.,  {Keppler} M.,  {Trifonov} T.,  {Penzlin} A.
  B.~T.,  {Avenhaus} H.,  {Henning} T.,   {Montesinos} M.,  2020, arXiv
  e-prints, \href {https://ui.adsabs.harvard.edu/abs/2020arXiv200711565B} {p.
  arXiv:2007.11565}

\bibitem[\protect\citeauthoryear{{Biller} et~al.,}{{Biller}
  et~al.}{2012}]{Biller2012}
{Biller} B.,  et~al., 2012, \mn@doi [\apjl] {10.1088/2041-8205/753/2/L38},
  \href {http://adsabs.harvard.edu/abs/2012ApJ...753L..38B} {753, L38}

\bibitem[\protect\citeauthoryear{{Birnstiel}, {Dullemond}  \&
  {Pinilla}}{{Birnstiel} et~al.}{2013}]{Birnstiel+2013}
{Birnstiel} T.,  {Dullemond} C.~P.,   {Pinilla} P.,  2013, \mn@doi [\aap]
  {10.1051/0004-6361/201220847}, \href
  {https://ui.adsabs.harvard.edu/abs/2013A&A...550L...8B} {550, L8}

\bibitem[\protect\citeauthoryear{{Calcino}, {Price}, {Pinte}, {van der Marel},
  {Ragusa}, {Dipierro}, {Cuello}  \& {Christiaens}}{{Calcino}
  et~al.}{2019}]{Calcino+2019}
{Calcino} J.,  {Price} D.~J.,  {Pinte} C.,  {van der Marel} N.,  {Ragusa} E.,
  {Dipierro} G.,  {Cuello} N.,   {Christiaens} V.,  2019, \mn@doi [\mnras]
  {10.1093/mnras/stz2770}, \href
  {https://ui.adsabs.harvard.edu/abs/2019MNRAS.490.2579C} {490, 2579}

\bibitem[\protect\citeauthoryear{{Calcino}, {Christiaens}, {Price}, {Pinte},
  {Davis}, {van der Marel}  \& {Cuello}}{{Calcino} et~al.}{2020}]{Calcino+2020}
{Calcino} J.,  {Christiaens} V.,  {Price} D.~J.,  {Pinte} C.,  {Davis} T.~M.,
  {van der Marel} N.,   {Cuello} N.,  2020, \mn@doi [\mnras]
  {10.1093/mnras/staa2468}, \href
  {https://ui.adsabs.harvard.edu/abs/2020MNRAS.498..639C} {498, 639}

\bibitem[\protect\citeauthoryear{{Calvet}, {D'Alessio}, {Hartmann}, {Wilner},
  {Walsh}  \& {Sitko}}{{Calvet} et~al.}{2002}]{Calvet+2002}
{Calvet} N.,  {D'Alessio} P.,  {Hartmann} L.,  {Wilner} D.,  {Walsh} A.,
  {Sitko} M.,  2002, \mn@doi [\apj] {10.1086/339061}, \href
  {https://ui.adsabs.harvard.edu/abs/2002ApJ...568.1008C} {568, 1008}

\bibitem[\protect\citeauthoryear{{Carney} et~al.,}{{Carney}
  et~al.}{2018}]{Carney+2018}
{Carney} M.~T.,  et~al., 2018, \mn@doi [\aap] {10.1051/0004-6361/201732384},
  \href {https://ui.adsabs.harvard.edu/abs/2018A&A...614A.106C} {614, A106}

\bibitem[\protect\citeauthoryear{{Casassus}}{{Casassus}}{2016}]{Casassus2016}
{Casassus} S.,  2016, \mn@doi [\pasa] {10.1017/pasa.2016.7}, \href
  {https://ui.adsabs.harvard.edu/abs/2016PASA...33...13C} {33, e013}

\bibitem[\protect\citeauthoryear{{Chen}, {K{\'o}sp{\'a}l}, {{\'A}brah{\'a}m},
  {Kreplin}, {Matter}  \& {Weigelt}}{{Chen} et~al.}{2018}]{Chen+2018}
{Chen} L.,  {K{\'o}sp{\'a}l} {\'A}.,  {{\'A}brah{\'a}m} P.,  {Kreplin} A.,
  {Matter} A.,   {Weigelt} G.,  2018, \mn@doi [\aap]
  {10.1051/0004-6361/201731627}, \href
  {https://ui.adsabs.harvard.edu/abs/2018A&A...609A..45C} {609, A45}

\bibitem[\protect\citeauthoryear{{Cuello}, {Gonzalez}  \& {Pignatale}}{{Cuello}
  et~al.}{2016}]{Cuello+2016}
{Cuello} N.,  {Gonzalez} J.~F.,   {Pignatale} F.~C.,  2016, \mn@doi [\mnras]
  {10.1093/mnras/stw396}, \href
  {https://ui.adsabs.harvard.edu/abs/2016MNRAS.458.2140C} {458, 2140}

\bibitem[\protect\citeauthoryear{{Dent}, {Torrelles}, {Osorio}, {Calvet}  \&
  {Anglada}}{{Dent} et~al.}{2006}]{Dent2006}
{Dent} W.~R.~F.,  {Torrelles} J.~M.,  {Osorio} M.,  {Calvet} N.,   {Anglada}
  G.,  2006, \mn@doi [\mnras] {10.1111/j.1365-2966.2005.09825.x}, \href
  {https://ui.adsabs.harvard.edu/abs/2006MNRAS.365.1283D} {365, 1283}

\bibitem[\protect\citeauthoryear{{Dong}, {Zhu}, {Rafikov}  \& {Stone}}{{Dong}
  et~al.}{2015}]{Dong+2015}
{Dong} R.,  {Zhu} Z.,  {Rafikov} R.~R.,   {Stone} J.~M.,  2015, \mn@doi [\apjl]
  {10.1088/2041-8205/809/1/L5}, \href
  {https://ui.adsabs.harvard.edu/abs/2015ApJ...809L...5D} {809, L5}

\bibitem[\protect\citeauthoryear{{Dullemond}, {K{\"u}ffmeier}, {Goicovic},
  {Fukagawa}, {Oehl}  \& {Kramer}}{{Dullemond} et~al.}{2019}]{Dullemond+2019}
{Dullemond} C.~P.,  {K{\"u}ffmeier} M.,  {Goicovic} F.,  {Fukagawa} M.,  {Oehl}
  V.,   {Kramer} M.,  2019, \mn@doi [\aap] {10.1051/0004-6361/201832632}, \href
  {https://ui.adsabs.harvard.edu/abs/2019A&A...628A..20D} {628, A20}

\bibitem[\protect\citeauthoryear{{Dunhill}, {Cuadra}  \& {Dougados}}{{Dunhill}
  et~al.}{2015}]{Dunhill+2015}
{Dunhill} A.~C.,  {Cuadra} J.,   {Dougados} C.,  2015, \mn@doi [\mnras]
  {10.1093/mnras/stv284}, \href
  {https://ui.adsabs.harvard.edu/abs/2015MNRAS.448.3545D} {448, 3545}

\bibitem[\protect\citeauthoryear{{Espaillat} et~al.,}{{Espaillat}
  et~al.}{2010}]{Espaillat+2010}
{Espaillat} C.,  et~al., 2010, \mn@doi [\apj] {10.1088/0004-637X/717/1/441},
  \href {https://ui.adsabs.harvard.edu/abs/2010ApJ...717..441E} {717, 441}

\bibitem[\protect\citeauthoryear{{Fedele} et~al.,}{{Fedele}
  et~al.}{2017}]{Fedele+2017}
{Fedele} D.,  et~al., 2017, \mn@doi [\aap] {10.1051/0004-6361/201629860}, \href
  {https://ui.adsabs.harvard.edu/abs/2017A&A...600A..72F} {600, A72}

\bibitem[\protect\citeauthoryear{{Flock}, {Ruge}, {Dzyurkevich}, {Henning},
  {Klahr}  \& {Wolf}}{{Flock} et~al.}{2015}]{Flock+2015}
{Flock} M.,  {Ruge} J.~P.,  {Dzyurkevich} N.,  {Henning} T.,  {Klahr} H.,
  {Wolf} S.,  2015, \mn@doi [\aap] {10.1051/0004-6361/201424693}, \href
  {https://ui.adsabs.harvard.edu/abs/2015A&A...574A..68F} {574, A68}

\bibitem[\protect\citeauthoryear{{Francis} \& {van der Marel}}{{Francis} \&
  {van der Marel}}{2020}]{Francis&Marel2020}
{Francis} L.,  {van der Marel} N.,  2020, \mn@doi [\apj]
  {10.3847/1538-4357/ab7b63}, \href
  {https://ui.adsabs.harvard.edu/abs/2020ApJ...892..111F} {892, 111}

\bibitem[\protect\citeauthoryear{{Fuente} et~al.,}{{Fuente}
  et~al.}{2017}]{Fuente+2017}
{Fuente} A.,  et~al., 2017, \mn@doi [\apjl] {10.3847/2041-8213/aa8558}, \href
  {https://ui.adsabs.harvard.edu/abs/2017ApJ...846L...3F} {846, L3}

\bibitem[\protect\citeauthoryear{{Fukagawa} et~al.,}{{Fukagawa}
  et~al.}{2004}]{Fukagawa+2004}
{Fukagawa} M.,  et~al., 2004, \mn@doi [\apjl] {10.1086/420699}, \href
  {https://ui.adsabs.harvard.edu/abs/2004ApJ...605L..53F} {605, L53}

\bibitem[\protect\citeauthoryear{{Gaia Collaboration} et~al.,}{{Gaia
  Collaboration} et~al.}{2018}]{Gaia2018}
{Gaia Collaboration} et~al., 2018, \mn@doi [\aap]
  {10.1051/0004-6361/201833051}, \href
  {https://ui.adsabs.harvard.edu/\#abs/2018A&A...616A...1G} {616, A1}

\bibitem[\protect\citeauthoryear{{Gaia Collaboration} et~al.,}{{Gaia
  Collaboration} et~al.}{2021}]{Gaia+2021}
{Gaia Collaboration} et~al., 2021, \mn@doi [\aap]
  {10.1051/0004-6361/202039657}, \href
  {https://ui.adsabs.harvard.edu/abs/2021A&A...649A...1G} {649, A1}

\bibitem[\protect\citeauthoryear{{Gonzalez}, {Laibe}  \& {Maddison}}{{Gonzalez}
  et~al.}{2017}]{Gonzalez+2017}
{Gonzalez} J.~F.,  {Laibe} G.,   {Maddison} S.~T.,  2017, \mn@doi [\mnras]
  {10.1093/mnras/stx016}, \href
  {https://ui.adsabs.harvard.edu/abs/2017MNRAS.467.1984G} {467, 1984}

\bibitem[\protect\citeauthoryear{{Grady} et~al.,}{{Grady}
  et~al.}{2007}]{Grady+2007}
{Grady} C.~A.,  et~al., 2007, \mn@doi [\apj] {10.1086/519757}, \href
  {https://ui.adsabs.harvard.edu/abs/2007ApJ...665.1391G} {665, 1391}

\bibitem[\protect\citeauthoryear{{Gratton} et~al.,}{{Gratton}
  et~al.}{2019}]{Gratton+2019}
{Gratton} R.,  et~al., 2019, \mn@doi [\aap] {10.1051/0004-6361/201834760},
  \href {https://ui.adsabs.harvard.edu/abs/2019A&A...623A.140G} {623, A140}

\bibitem[\protect\citeauthoryear{{Gray}, {Riggs}, {Koen}, {Murphy}, {Newsome},
  {Corbally}, {Cheng}  \& {Neff}}{{Gray} et~al.}{2017}]{Gray+2017}
{Gray} R.~O.,  {Riggs} Q.~S.,  {Koen} C.,  {Murphy} S.~J.,  {Newsome} I.~M.,
  {Corbally} C.~J.,  {Cheng} K.~P.,   {Neff} J.~E.,  2017, \mn@doi [\aj]
  {10.3847/1538-3881/aa6d5e}, \href
  {https://ui.adsabs.harvard.edu/abs/2017AJ....154...31G} {154, 31}

\bibitem[\protect\citeauthoryear{{Haffert}, {Bohn}, {de Boer}, {Snellen},
  {Brinchmann}, {Girard}, {Keller}  \& {Bacon}}{{Haffert}
  et~al.}{2019}]{Haffert2019}
{Haffert} S.~Y.,  {Bohn} A.~J.,  {de Boer} J.,  {Snellen} I.~A.~G.,
  {Brinchmann} J.,  {Girard} J.~H.,  {Keller} C.~U.,   {Bacon} R.,  2019,
  \mn@doi [Nature Astronomy] {10.1038/s41550-019-0780-5}, \href
  {https://ui.adsabs.harvard.edu/abs/2019NatAs.tmp..329H} {p.~329}

\bibitem[\protect\citeauthoryear{{Hashimoto} et~al.,}{{Hashimoto}
  et~al.}{2011}]{Hashimoto+2011}
{Hashimoto} J.,  et~al., 2011, \mn@doi [\apjl] {10.1088/2041-8205/729/2/L17},
  \href {https://ui.adsabs.harvard.edu/abs/2011ApJ...729L..17H} {729, L17}

\bibitem[\protect\citeauthoryear{{Hirsh}, {Price}, {Gonzalez},
  {Ubeira-Gabellini}  \& {Ragusa}}{{Hirsh} et~al.}{2020}]{Hirsh+2020}
{Hirsh} K.,  {Price} D.~J.,  {Gonzalez} J.-F.,  {Ubeira-Gabellini} M.~G.,
  {Ragusa} E.,  2020, \mn@doi [\mnras] {10.1093/mnras/staa2536}, \href
  {https://ui.adsabs.harvard.edu/abs/2020MNRAS.498.2936H} {498, 2936}

\bibitem[\protect\citeauthoryear{{Ireland} \& {Kraus}}{{Ireland} \&
  {Kraus}}{2008}]{Ireland&Kraus2008}
{Ireland} M.~J.,  {Kraus} A.~L.,  2008, \mn@doi [\apjl] {10.1086/588216}, \href
  {https://ui.adsabs.harvard.edu/abs/2008ApJ...678L..59I} {678, L59}

\bibitem[\protect\citeauthoryear{{Keppler} et~al.,}{{Keppler}
  et~al.}{2018}]{Keppler+2018}
{Keppler} M.,  et~al., 2018, \mn@doi [\aap] {10.1051/0004-6361/201832957},
  \href {https://ui.adsabs.harvard.edu/abs/2018A&A...617A..44K} {617, A44}

\bibitem[\protect\citeauthoryear{{Kuffmeier}, {Goicovic}  \&
  {Dullemond}}{{Kuffmeier} et~al.}{2020}]{Kuffmeier+2020}
{Kuffmeier} M.,  {Goicovic} F.~G.,   {Dullemond} C.~P.,  2020, \mn@doi [\aap]
  {10.1051/0004-6361/201936820}, \href
  {https://ui.adsabs.harvard.edu/abs/2020A&A...633A...3K} {633, A3}

\bibitem[\protect\citeauthoryear{{Lacour} et~al.,}{{Lacour}
  et~al.}{2016}]{Lacour+2016}
{Lacour} S.,  et~al., 2016, \mn@doi [\aap] {10.1051/0004-6361/201527863}, \href
  {https://ui.adsabs.harvard.edu/abs/2016A&A...590A..90L} {590, A90}

\bibitem[\protect\citeauthoryear{{Laibe} \& {Price}}{{Laibe} \&
  {Price}}{2012a}]{Laibe12a}
{Laibe} G.,  {Price} D.~J.,  2012a, \mn@doi [\mnras]
  {10.1111/j.1365-2966.2011.20202.x}, \href
  {https://ui.adsabs.harvard.edu/#abs/2012MNRAS.420.2345L} {420, 2345}

\bibitem[\protect\citeauthoryear{{Laibe} \& {Price}}{{Laibe} \&
  {Price}}{2012b}]{Laibe12b}
{Laibe} G.,  {Price} D.~J.,  2012b, \mn@doi [\mnras]
  {10.1111/j.1365-2966.2011.20201.x}, \href
  {https://ui.adsabs.harvard.edu/#abs/2012MNRAS.420.2365L} {420, 2365}

\bibitem[\protect\citeauthoryear{{Lazareff} et~al.,}{{Lazareff}
  et~al.}{2017}]{Lazareff+2017}
{Lazareff} B.,  et~al., 2017, \mn@doi [\aap] {10.1051/0004-6361/201629305},
  \href {https://ui.adsabs.harvard.edu/abs/2017A&A...599A..85L} {599, A85}

\bibitem[\protect\citeauthoryear{{Lesur}, {Kunz}  \& {Fromang}}{{Lesur}
  et~al.}{2014}]{Lesur+2014}
{Lesur} G.,  {Kunz} M.~W.,   {Fromang} S.,  2014, \mn@doi [\aap]
  {10.1051/0004-6361/201423660}, \href
  {https://ui.adsabs.harvard.edu/abs/2014A&A...566A..56L} {566, A56}

\bibitem[\protect\citeauthoryear{{Li} \& {Draine}}{{Li} \&
  {Draine}}{2001}]{Li&Draine2001}
{Li} A.,  {Draine} B.~T.,  2001, \mn@doi [\apj] {10.1086/323147}, \href
  {https://ui.adsabs.harvard.edu/abs/2001ApJ...554..778L} {554, 778}

\bibitem[\protect\citeauthoryear{{Ligi} et~al.,}{{Ligi}
  et~al.}{2018}]{Ligi+2018}
{Ligi} R.,  et~al., 2018, \mn@doi [\mnras] {10.1093/mnras/stx2318}, \href
  {https://ui.adsabs.harvard.edu/abs/2018MNRAS.473.1774L} {473, 1774}

\bibitem[\protect\citeauthoryear{{Lin} \& {Papaloizou}}{{Lin} \&
  {Papaloizou}}{1986}]{Lin+1986}
{Lin} D.~N.~C.,  {Papaloizou} J.,  1986, \mn@doi [\apj] {10.1086/164653}, \href
  {https://ui.adsabs.harvard.edu/abs/1986ApJ...309..846L} {309, 846}

\bibitem[\protect\citeauthoryear{{Lodato} \& {Price}}{{Lodato} \&
  {Price}}{2010}]{Lodato&Price2010}
{Lodato} G.,  {Price} D.~J.,  2010, \mn@doi [\mnras]
  {10.1111/j.1365-2966.2010.16526.x}, \href
  {http://adsabs.harvard.edu/abs/2010MNRAS.405.1212L} {405, 1212}

\bibitem[\protect\citeauthoryear{{Lyra} \& {Lin}}{{Lyra} \&
  {Lin}}{2013}]{Lyra&Lin2013}
{Lyra} W.,  {Lin} M.-K.,  2013, \mn@doi [\apj] {10.1088/0004-637X/775/1/17},
  \href {https://ui.adsabs.harvard.edu/abs/2013ApJ...775...17L} {775, 17}

\bibitem[\protect\citeauthoryear{{Mac{\'\i}as}, {Anglada}, {Osorio},
  {Torrelles}, {Carrasco-Gonz{\'a}lez}, {G{\'o}mez}, {Rodr{\'\i}guez}  \&
  {Sierra}}{{Mac{\'\i}as} et~al.}{2017}]{Macias+2017}
{Mac{\'\i}as} E.,  {Anglada} G.,  {Osorio} M.,  {Torrelles} J.~M.,
  {Carrasco-Gonz{\'a}lez} C.,  {G{\'o}mez} J.~F.,  {Rodr{\'\i}guez} L.~F.,
  {Sierra} A.,  2017, \mn@doi [\apj] {10.3847/1538-4357/aa6620}, \href
  {https://ui.adsabs.harvard.edu/abs/2017ApJ...838...97M} {838, 97}

\bibitem[\protect\citeauthoryear{{Mac{\'\i}as} et~al.,}{{Mac{\'\i}as}
  et~al.}{2019}]{Macias+2019}
{Mac{\'\i}as} E.,  et~al., 2019, \mn@doi [\apj] {10.3847/1538-4357/ab31a2},
  \href {https://ui.adsabs.harvard.edu/abs/2019ApJ...881..159M} {881, 159}

\bibitem[\protect\citeauthoryear{{Meeus} et~al.,}{{Meeus}
  et~al.}{2010}]{Meeus2010}
{Meeus} G.,  et~al., 2010, \mn@doi [\aap] {10.1051/0004-6361/201014557}, \href
  {https://ui.adsabs.harvard.edu/abs/2010A&A...518L.124M} {518, L124}

\bibitem[\protect\citeauthoryear{{Monnier} et~al.,}{{Monnier}
  et~al.}{2017}]{Monnier+2017}
{Monnier} J.~D.,  et~al., 2017, \mn@doi [\apj] {10.3847/1538-4357/aa6248},
  \href {https://ui.adsabs.harvard.edu/abs/2017ApJ...838...20M} {838, 20}

\bibitem[\protect\citeauthoryear{{Montesinos}, {Garrido-Deutelmoser},
  {Olofsson}, {Giuppone}, {Cuadra}, {Bayo}, {Sucerquia}  \&
  {Cuello}}{{Montesinos} et~al.}{2020}]{Montesinos+2020}
{Montesinos} M.,  {Garrido-Deutelmoser} J.,  {Olofsson} J.,  {Giuppone} C.~A.,
  {Cuadra} J.,  {Bayo} A.,  {Sucerquia} M.,   {Cuello} N.,  2020, \mn@doi
  [\aap] {10.1051/0004-6361/202038758}, \href
  {https://ui.adsabs.harvard.edu/abs/2020A&A...642A.224M} {642, A224}

\bibitem[\protect\citeauthoryear{{Montesinos}, {Cuello}, {Olofsson}, {Cuadra},
  {Bayo}, {Bertrang}  \& {Perrot}}{{Montesinos} et~al.}{2021}]{Montesinos+2021}
{Montesinos} M.,  {Cuello} N.,  {Olofsson} J.,  {Cuadra} J.,  {Bayo} A.,
  {Bertrang} G. H.~M.,   {Perrot} C.,  2021, \mn@doi [\apj]
  {10.3847/1538-4357/abe3fc}, \href
  {https://ui.adsabs.harvard.edu/abs/2021ApJ...910...31M} {910, 31}

\bibitem[\protect\citeauthoryear{{Nealon}, {Dipierro}, {Alexander}, {Martin}
  \& {Nixon}}{{Nealon} et~al.}{2018}]{Nealon+2018}
{Nealon} R.,  {Dipierro} G.,  {Alexander} R.,  {Martin} R.~G.,   {Nixon} C.,
  2018, \mn@doi [\mnras] {10.1093/mnras/sty2267}, \href
  {https://ui.adsabs.harvard.edu/abs/2018MNRAS.481...20N} {481, 20}

\bibitem[\protect\citeauthoryear{{Osorio} et~al.,}{{Osorio}
  et~al.}{2014}]{Osorio+2014}
{Osorio} M.,  et~al., 2014, \mn@doi [\apjl] {10.1088/2041-8205/791/2/L36},
  \href {https://ui.adsabs.harvard.edu/abs/2014ApJ...791L..36O} {791, L36}

\bibitem[\protect\citeauthoryear{{Pacheco-V{\'a}zquez}
  et~al.,}{{Pacheco-V{\'a}zquez} et~al.}{2016}]{Pacheco-Vazquez+2016}
{Pacheco-V{\'a}zquez} S.,  et~al., 2016, \mn@doi [\aap]
  {10.1051/0004-6361/201527089}, \href
  {https://ui.adsabs.harvard.edu/abs/2016A&A...589A..60P} {589, A60}

\bibitem[\protect\citeauthoryear{{Pani{\'c}}, {Hogerheijde}, {Wilner}  \&
  {Qi}}{{Pani{\'c}} et~al.}{2008}]{Panic+2008}
{Pani{\'c}} O.,  {Hogerheijde} M.~R.,  {Wilner} D.,   {Qi} C.,  2008, \mn@doi
  [\aap] {10.1051/0004-6361:20079261}, \href
  {https://ui.adsabs.harvard.edu/abs/2008A&A...491..219P} {491, 219}

\bibitem[\protect\citeauthoryear{{P{\'e}rez}, {Isella}, {Carpenter}  \&
  {Chandler}}{{P{\'e}rez} et~al.}{2014}]{Perez+2014}
{P{\'e}rez} L.~M.,  {Isella} A.,  {Carpenter} J.~M.,   {Chandler} C.~J.,  2014,
  \mn@doi [\apjl] {10.1088/2041-8205/783/1/L13}, \href
  {https://ui.adsabs.harvard.edu/abs/2014ApJ...783L..13P} {783, L13}

\bibitem[\protect\citeauthoryear{{P{\'e}rez}, {Casassus}, {Baruteau}, {Dong},
  {Hales}  \& {Cieza}}{{P{\'e}rez} et~al.}{2019}]{Perez+2019}
{P{\'e}rez} S.,  {Casassus} S.,  {Baruteau} C.,  {Dong} R.,  {Hales} A.,
  {Cieza} L.,  2019, \mn@doi [\aj] {10.3847/1538-3881/ab1f88}, \href
  {https://ui.adsabs.harvard.edu/abs/2019AJ....158...15P} {158, 15}

\bibitem[\protect\citeauthoryear{{Pinilla}, {Birnstiel}, {Ricci}, {Dullemond},
  {Uribe}, {Testi}  \& {Natta}}{{Pinilla} et~al.}{2012a}]{Pinilla+2012a}
{Pinilla} P.,  {Birnstiel} T.,  {Ricci} L.,  {Dullemond} C.~P.,  {Uribe} A.~L.,
   {Testi} L.,   {Natta} A.,  2012a, \mn@doi [\aap]
  {10.1051/0004-6361/201118204}, \href
  {https://ui.adsabs.harvard.edu/abs/2012A&A...538A.114P} {538, A114}

\bibitem[\protect\citeauthoryear{{Pinilla}, {Benisty}  \&
  {Birnstiel}}{{Pinilla} et~al.}{2012b}]{Pinilla+2012b}
{Pinilla} P.,  {Benisty} M.,   {Birnstiel} T.,  2012b, \mn@doi [\aap]
  {10.1051/0004-6361/201219315}, \href
  {https://ui.adsabs.harvard.edu/abs/2012A&A...545A..81P} {545, A81}

\bibitem[\protect\citeauthoryear{{Pinilla}, {Flock}, {Ovelar}  \&
  {Birnstiel}}{{Pinilla} et~al.}{2016}]{Pinilla+2016}
{Pinilla} P.,  {Flock} M.,  {Ovelar} M. d.~J.,   {Birnstiel} T.,  2016, \mn@doi
  [\aap] {10.1051/0004-6361/201628441}, \href
  {https://ui.adsabs.harvard.edu/abs/2016A&A...596A..81P} {596, A81}

\bibitem[\protect\citeauthoryear{{Pinilla}, {Pohl}, {Stammler}  \&
  {Birnstiel}}{{Pinilla} et~al.}{2017}]{Pinilla+2017}
{Pinilla} P.,  {Pohl} A.,  {Stammler} S.~M.,   {Birnstiel} T.,  2017, \mn@doi
  [\apj] {10.3847/1538-4357/aa7edb}, \href
  {https://ui.adsabs.harvard.edu/abs/2017ApJ...845...68P} {845, 68}

\bibitem[\protect\citeauthoryear{{Pinte}, {M{\'e}nard}, {Duch{\^e}ne}  \&
  {Bastien}}{{Pinte} et~al.}{2006}]{pinte2006}
{Pinte} C.,  {M{\'e}nard} F.,  {Duch{\^e}ne} G.,   {Bastien} P.,  2006, \mn@doi
  [\aap] {10.1051/0004-6361:20053275}, \href
  {http://adsabs.harvard.edu/abs/2006A%26A...459..797P} {459, 797}

\bibitem[\protect\citeauthoryear{{Pinte}, {Harries}, {Min}, {Watson},
  {Dullemond}, {Woitke}, {M{\'e}nard}  \& {Dur{\'a}n-Rojas}}{{Pinte}
  et~al.}{2009}]{pinte2009}
{Pinte} C.,  {Harries} T.~J.,  {Min} M.,  {Watson} A.~M.,  {Dullemond} C.~P.,
  {Woitke} P.,  {M{\'e}nard} F.,   {Dur{\'a}n-Rojas} M.~C.,  2009, \mn@doi
  [\aap] {10.1051/0004-6361/200811555}, \href
  {http://adsabs.harvard.edu/abs/2009A%26A...498..967P} {498, 967}

\bibitem[\protect\citeauthoryear{{Poblete}, {Cuello}  \& {Cuadra}}{{Poblete}
  et~al.}{2019}]{Poblete+2019}
{Poblete} P.~P.,  {Cuello} N.,   {Cuadra} J.,  2019, \mn@doi [\mnras]
  {10.1093/mnras/stz2297}, \href
  {https://ui.adsabs.harvard.edu/abs/2019MNRAS.489.2204P} {489, 2204}

\bibitem[\protect\citeauthoryear{{Poblete}, {Calcino}, {Cuello}, {Mac{\'\i}as},
  {Ribas}, {Price}, {Cuadra}  \& {Pinte}}{{Poblete}
  et~al.}{2020}]{Poblete+2020}
{Poblete} P.~P.,  {Calcino} J.,  {Cuello} N.,  {Mac{\'\i}as} E.,  {Ribas}
  {\'A}.,  {Price} D.~J.,  {Cuadra} J.,   {Pinte} C.,  2020, \mn@doi [\mnras]
  {10.1093/mnras/staa1655}, \href
  {https://ui.adsabs.harvard.edu/abs/2020MNRAS.496.2362P} {496, 2362}

\bibitem[\protect\citeauthoryear{{Pohl} et~al.,}{{Pohl}
  et~al.}{2017}]{Pohl+2017}
{Pohl} A.,  et~al., 2017, \mn@doi [\apj] {10.3847/1538-4357/aa94c2}, \href
  {https://ui.adsabs.harvard.edu/abs/2017ApJ...850...52P} {850, 52}

\bibitem[\protect\citeauthoryear{{Price} \& {Laibe}}{{Price} \&
  {Laibe}}{2020}]{Price&Laibe+2020}
{Price} D.~J.,  {Laibe} G.,  2020, \mn@doi [\mnras] {10.1093/mnras/staa1366},
  \href {https://ui.adsabs.harvard.edu/abs/2020MNRAS.495.3929P} {495, 3929}

\bibitem[\protect\citeauthoryear{{Price} et~al.,}{{Price}
  et~al.}{2018a}]{PricePH+2018b}
{Price} D.~J.,  et~al., 2018a, \mn@doi [Publications of the Astronomical
  Society of Australia] {10.1017/pasa.2018.25}, \href
  {https://ui.adsabs.harvard.edu/\#abs/2018PASA...35...31P} {35, e031}

\bibitem[\protect\citeauthoryear{{Price} et~al.,}{{Price}
  et~al.}{2018b}]{Price+2018}
{Price} D.~J.,  et~al., 2018b, \mn@doi [\mnras] {10.1093/mnras/sty647}, \href
  {https://ui.adsabs.harvard.edu/#abs/2018MNRAS.477.1270P} {477, 1270}

\bibitem[\protect\citeauthoryear{{Quanz}, {Avenhaus}, {Buenzli}, {Garufi},
  {Schmid}  \& {Wolf}}{{Quanz} et~al.}{2013}]{Quanz+2013}
{Quanz} S.~P.,  {Avenhaus} H.,  {Buenzli} E.,  {Garufi} A.,  {Schmid} H.~M.,
  {Wolf} S.,  2013, \mn@doi [\apjl] {10.1088/2041-8205/766/1/L2}, \href
  {https://ui.adsabs.harvard.edu/abs/2013ApJ...766L...2Q} {766, L2}

\bibitem[\protect\citeauthoryear{{Ragusa}, {Dipierro}, {Lodato}, {Laibe}  \&
  {Price}}{{Ragusa} et~al.}{2017}]{Ragusa+2017}
{Ragusa} E.,  {Dipierro} G.,  {Lodato} G.,  {Laibe} G.,   {Price} D.~J.,  2017,
  \mn@doi [\mnras] {10.1093/mnras/stw2456}, \href
  {https://ui.adsabs.harvard.edu/abs/2017MNRAS.464.1449R} {464, 1449}

\bibitem[\protect\citeauthoryear{{Ragusa}, {Alexander}, {Calcino}, {Hirsh}  \&
  {Price}}{{Ragusa} et~al.}{2020}]{Ragusa+2020}
{Ragusa} E.,  {Alexander} R.,  {Calcino} J.,  {Hirsh} K.,   {Price} D.~J.,
  2020, \mn@doi [\mnras] {10.1093/mnras/staa2954}, \href
  {https://ui.adsabs.harvard.edu/abs/2020MNRAS.499.3362R} {499, 3362}

\bibitem[\protect\citeauthoryear{{Ragusa} et~al.,}{{Ragusa}
  et~al.}{2021}]{Ragusa+2021}
{Ragusa} E.,  et~al., 2021, \mn@doi [\mnras] {10.1093/mnras/stab2179}, \href
  {https://ui.adsabs.harvard.edu/abs/2021MNRAS.507.1157R} {507, 1157}

\bibitem[\protect\citeauthoryear{{Raman}, {Lisanti}, {Wilner}, {Qi}  \&
  {Hogerheijde}}{{Raman} et~al.}{2006}]{Raman+2006}
{Raman} A.,  {Lisanti} M.,  {Wilner} D.~J.,  {Qi} C.,   {Hogerheijde} M.,
  2006, \mn@doi [\aj] {10.1086/500587}, \href
  {https://ui.adsabs.harvard.edu/abs/2006AJ....131.2290R} {131, 2290}

\bibitem[\protect\citeauthoryear{{Riols} \& {Lesur}}{{Riols} \&
  {Lesur}}{2018}]{Riols&Lesur2018}
{Riols} A.,  {Lesur} G.,  2018, \mn@doi [\aap] {10.1051/0004-6361/201833212},
  \href {https://ui.adsabs.harvard.edu/abs/2018A&A...617A.117R} {617, A117}

\bibitem[\protect\citeauthoryear{{Shakura} \& {Sunyaev}}{{Shakura} \&
  {Sunyaev}}{1973}]{Shakura&Sunyaev73}
{Shakura} N.~I.,  {Sunyaev} R.~A.,  1973, \aap, \href
  {http://cdsads.u-strasbg.fr/abs/1973A%26A....24..337S} {24, 337}

\bibitem[\protect\citeauthoryear{{Stolker} et~al.,}{{Stolker}
  et~al.}{2016}]{Stolker+2016}
{Stolker} T.,  et~al., 2016, \mn@doi [\aap] {10.1051/0004-6361/201528039},
  \href {https://ui.adsabs.harvard.edu/abs/2016A&A...595A.113S} {595, A113}

\bibitem[\protect\citeauthoryear{{Strom}, {Strom}, {Edwards}, {Cabrit}  \&
  {Skrutskie}}{{Strom} et~al.}{1989}]{Strom+1989}
{Strom} K.~M.,  {Strom} S.~E.,  {Edwards} S.,  {Cabrit} S.,   {Skrutskie}
  M.~F.,  1989, \mn@doi [\aj] {10.1086/115085}, \href
  {https://ui.adsabs.harvard.edu/abs/1989AJ.....97.1451S} {97, 1451}

\bibitem[\protect\citeauthoryear{{Suriano}, {Li}, {Krasnopolsky}  \&
  {Shang}}{{Suriano} et~al.}{2018}]{Suriano+2018}
{Suriano} S.~S.,  {Li} Z.-Y.,  {Krasnopolsky} R.,   {Shang} H.,  2018, \mn@doi
  [\mnras] {10.1093/mnras/sty717}, \href
  {https://ui.adsabs.harvard.edu/abs/2018MNRAS.477.1239S} {477, 1239}

\bibitem[\protect\citeauthoryear{{Toci}, {Lodato}, {Fedele}, {Testi}  \&
  {Pinte}}{{Toci} et~al.}{2020}]{Toci+2020}
{Toci} C.,  {Lodato} G.,  {Fedele} D.,  {Testi} L.,   {Pinte} C.,  2020,
  \mn@doi [\apjl] {10.3847/2041-8213/ab5c87}, \href
  {https://ui.adsabs.harvard.edu/abs/2020ApJ...888L...4T} {888, L4}

\bibitem[\protect\citeauthoryear{{Tokovinin}}{{Tokovinin}}{2018}]{Tokovinin2018}
{Tokovinin} A.,  2018, \mn@doi [\pasp] {10.1088/1538-3873/aaa7d9}, \href
  {https://ui.adsabs.harvard.edu/abs/2018PASP..130c5002T} {130, 035002}

\bibitem[\protect\citeauthoryear{{Wang} et~al.,}{{Wang}
  et~al.}{2020}]{Wang+2020}
{Wang} J.~J.,  et~al., 2020, \mn@doi [\aj] {10.3847/1538-3881/ab8aef}, \href
  {https://ui.adsabs.harvard.edu/abs/2020AJ....159..263W} {159, 263}

\bibitem[\protect\citeauthoryear{{Weingartner} \& {Draine}}{{Weingartner} \&
  {Draine}}{2001}]{weingartner2001}
{Weingartner} J.~C.,  {Draine} B.~T.,  2001, \mn@doi [\apj] {10.1086/318651},
  \href {https://ui.adsabs.harvard.edu/abs/2001ApJ...548..296W} {548, 296}

\bibitem[\protect\citeauthoryear{{Yu}, {Teague}, {Bae}  \& {{\"O}berg}}{{Yu}
  et~al.}{2021}]{Yu+2021}
{Yu} H.,  {Teague} R.,  {Bae} J.,   {{\"O}berg} K.,  2021, \mn@doi [\apjl]
  {10.3847/2041-8213/ac283e}, \href
  {https://ui.adsabs.harvard.edu/abs/2021ApJ...920L..33Y} {920, L33}

\bibitem[\protect\citeauthoryear{{Zhang} et~al.,}{{Zhang}
  et~al.}{2018}]{Zhang+2018}
{Zhang} S.,  et~al., 2018, \mn@doi [\apjl] {10.3847/2041-8213/aaf744}, \href
  {https://ui.adsabs.harvard.edu/abs/2018ApJ...869L..47Z} {869, L47}

\bibitem[\protect\citeauthoryear{{Zhu}, {Stone}, {Rafikov}  \& {Bai}}{{Zhu}
  et~al.}{2014}]{Zhu&Stone2014}
{Zhu} Z.,  {Stone} J.~M.,  {Rafikov} R.~R.,   {Bai} X.-n.,  2014, \mn@doi
  [\apj] {10.1088/0004-637X/785/2/122}, \href
  {https://ui.adsabs.harvard.edu/abs/2014ApJ...785..122Z} {785, 122}

\bibitem[\protect\citeauthoryear{{van der Marel} et~al.,}{{van der Marel}
  et~al.}{2013}]{vanderMarel+2013}
{van der Marel} N.,  et~al., 2013, \mn@doi [Science] {10.1126/science.1236770},
  \href {https://ui.adsabs.harvard.edu/\#abs/2013Sci...340.1199V} {340, 1199}

\bibitem[\protect\citeauthoryear{{van der Marel} et~al.,}{{van der Marel}
  et~al.}{2018}]{vanderMarel+2018}
{van der Marel} N.,  et~al., 2018, \mn@doi [\apj] {10.3847/1538-4357/aaaa6b},
  \href {https://ui.adsabs.harvard.edu/abs/2018ApJ...854..177V} {854, 177}

\bibitem[\protect\citeauthoryear{{van der Marel} et~al.,}{{van der Marel}
  et~al.}{2021}]{vanderMarel+2021}
{van der Marel} N.,  et~al., 2021, \mn@doi [\aj] {10.3847/1538-3881/abc3ba},
  \href {https://ui.adsabs.harvard.edu/abs/2021AJ....161...33V} {161, 33}

\makeatother
\end{thebibliography}

\appendix

\section{Numerical tests}

\begin{figure*}
\centering
\begin{center}
    \includegraphics[width=0.8\textwidth]{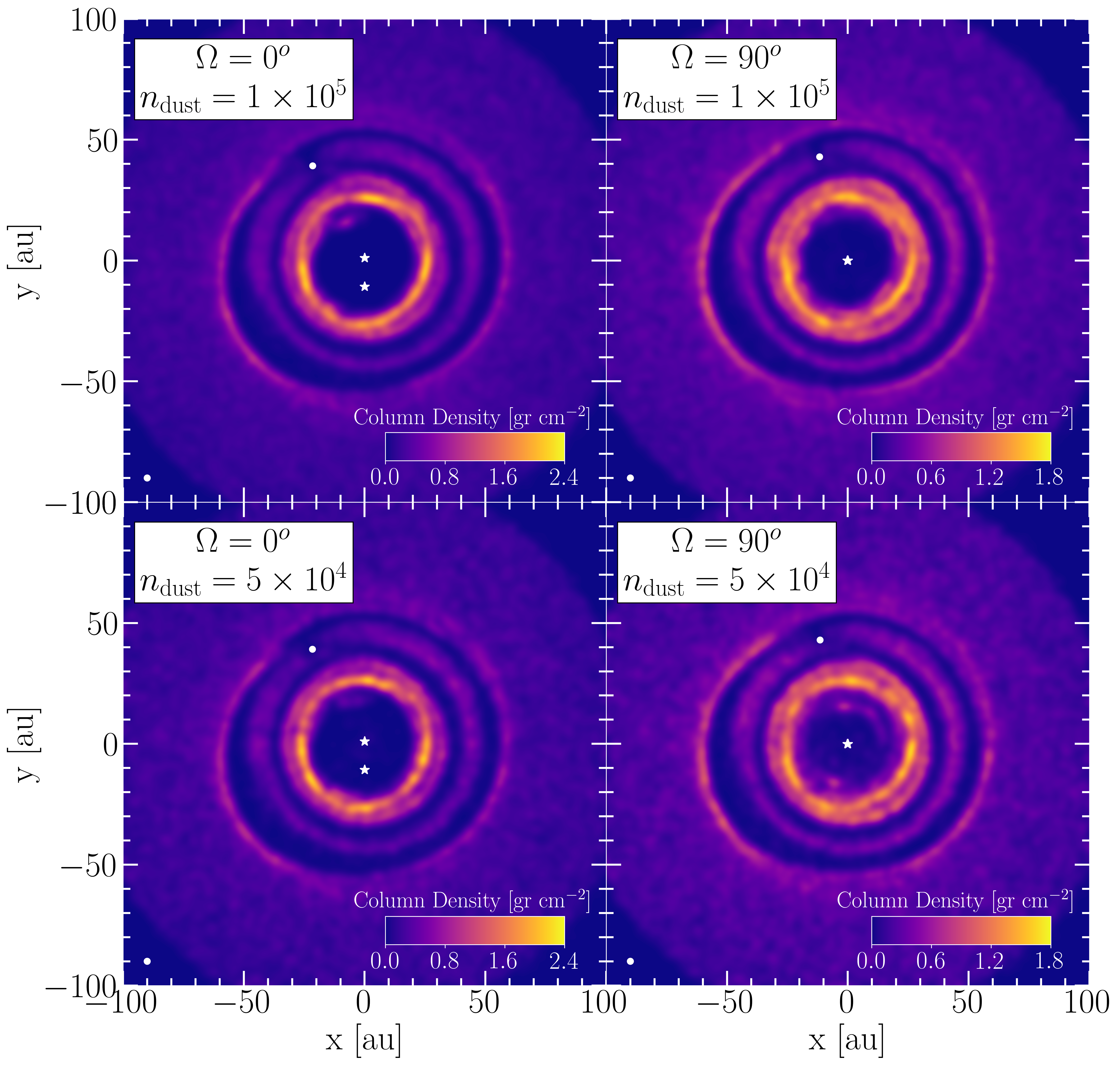} 
    \caption{ \ans{Dust morphology after 50 binary orbits for different binary configurations and dust particles resolutions. The first row shows the simulations with $1\times10^5$ dust particles for $\Omega=0\degree$ (left) and $\Omega=90\degree$ (right), whilst the bottom row shows the same simulations using $5\times10^4$ dust particles.} }
    \label{fig:A1}
\end{center}
\end{figure*}

\ans{Artificial dust clumping can appear when the dust/gas resolution is not optimal. In order to test it, we performed extra simulations with different resolutions for both binary configurations. Figure \ref{fig:A1} displays the dust morphology after 50 binary orbits for two dust particles resolution: $1\times10^5$ in the upper row, and $5\times10^4$ in the bottom row. The gas resolution is kept at $5\times10^5$. We observe that the resolution does not affect the formation of the dust structures, namely rings and clumps.
A similar test was presented in \citet{Poblete+2019}, but for a binary without a planet. Those results also show that the dust clumping is triggered by hydrodynamical effects rather than numerical ones.}

\label{lastpage}

\end{document}